\documentclass[showpacs,twocolumn,aps,pre]{revtex4}
\usepackage{graphicx}
\newcommand{\al}[0]{&\!\!}
\newcommand{\var}[0]{\bar{\varepsilon}}

\begin{document}

\title{Parametric instabilities in magnetized multicomponent plasmas}

\author{M. P. Hertzberg}
\author{N. F. Cramer}
\author{S. V. Vladimirov}
 \email{S.Vladimirov@physics.usyd.edu.au}
 \homepage{http://www.physics.usyd.edu.au/~vladimi}
\affiliation{School of Physics,
The University of Sydney, New South Wales 2006, Australia}


\begin{abstract}
This paper investigates the excitation of various natural modes in
a magnetized bi-ion or dusty plasma. The excitation is provided by
parametrically pumping the magnetic field. Here two ion-like
species are allowed to be fully mobile. This generalizes our previous work
where the second heavy species was taken to be stationary. Their
collection of charge from the background neutral plasma modifies
the dispersion properties of the pump and excited waves. The
introduction of an extra mobile species adds extra modes to both
these types of waves. We firstly investigate the pump wave in
detail, in the case where the background magnetic field is
perpendicular to the direction of propagation of the pump wave.
Then we derive the dispersion equation relating the pump to the
excited wave for modes propagating parallel to the background magnetic
field. It is found that there are a total of twelve resonant interactions
allowed, whose various growth rates are calculated and discussed.
\end{abstract}


\maketitle

\section{Introduction}
The basic natural modes of magnetized plasmas such as those that
occur in molecular clouds, cometary plasmas and stellar
atmospheres are of great interest. When the frequencies are low
then a class of linear waves, referred to collectively as Alfv\'en waves,
are known to exist and are of importance to the understanding of
many basic plasma phenomena \cite{Movspitz,Movpil}. The linear approximation to these waves breaks
down at large amplitudes where
nonlinear effects become important in their propagation. One such
large amplitude wave is a
magnetoacoustic wave, which modifies the background magnetic
field in an oscillatory fashion, and so can be considered as a pump wave that drives other waves nonlinearly.
Such large-amplitude pump waves may occur in conditions such as seen in solar
and space plasmas. For example, solar shock waves can set up large amplitude
standing magnetoacoustic waves in coronal loops or magnetic flux tubes
\cite{MovSakai1,MovSakai2,Movdurrant}.

By considering perturbations of this
large amplitude pump, we can investigate the possibility of
exciting natural modes of the system, such as Alfv\'en waves, due to a resonant interaction
between the pump wave and the natural waves.
In the single ion species case this basic phenomenon was
predicted by \cite{MovMont,MovVahal}, and
subsequently pursued by several authors \cite{MovCramer,MovMjol,MovMoral}.
For a pump magnetic field parallel to the background magnetic field in a single ion species plasma, the excited waves 
are Alfv\'en waves travelling in opposite directions along the magnetic field \cite{MovMont,MovVahal}.
If ion-cyclotron effects are included, the excited waves are
the fast and slow (ion--cyclotron) Alfv\'en waves travelling in opposite directions \cite{MovCramer}.
More general behavior is allowed in the case where the excited waves are
permitted to travel obliquely to the magnetic field \cite{MovCramer76}.

The parametric excitation of waves in a dusty magnetized plasma has been investigated in Ref.\,\cite{MovHertz}, but in the
approximation where the dust is taken to be immobile. In that case the oppositely travelling pair of waves are 
modified, with the presence of a cutoff frequency
in the fast Alfv\'en wave. Furthermore, an interaction between a pair of fast waves or a pair
of slow waves was found to exist. The corresponding growth rates of the slow-fast and fast-fast pairs
were maximized as a function of the dust concentration.

Many space and laboratory plasmas are multicomponent, i.e.,
contain multiple ion species. It is therefore of great interest to
investigate the effects of the additional species on the linear and nonlinear properties of the waves in the plasma. 
For instance, it is often the case that in the presence
of an additional ion component an extra
mode is excited, or forbidden regions of frequency are introduced.
One area of study of multi-ion plasmas is the bi-ion plasma.
The inclusion of one extra ion species often captures the basic
information of several extra species. The bi-ion plasma has particular importance in plasma
fusion, laboratory plasmas and in astrophysical environments, with
the secondary ion usually positively charged.

A dusty plasma adds another level of interest to the topic of multicomponent plasmas,
due to the dust properties. Dust is an additional
impurity of large mass and often of negative charge. Under the simplest approximation, all the dust grains may be 
considered to have the same mass and equilibrium charge, and so are equivalent to a second ion species. This is the 
case considered in this paper.
It is known that the inclusion of dust in a plasma
may introduce cutoff frequencies into the basic
Alfv\'en waves, and introduce a low frequency mode,
whose nature is different from that in a
bi-ion plasma due to the dust grain's extremely high mass \cite{MovVlad,MovCrambk}. Qualitative differences
to the bi-ion case arise when charge perturbations of the grains are included, leading to an additional damping 
mechanism \cite{Movvlprop}, or when a
spectrum of dust grain sizes and charges is allowed for \cite{Movspectrum}.
Dust is found, in varying amounts, in many astrophysical and space
environments such as molecular clouds and the rings of Saturn.
Dusty plasmas have also been studied closely experimentally,
since the heavy grain mass introduces low frequency effects
that may be studied in real time.

An immediate consequence of the presence of an additional ion
species is its modification of the background free-electron number density.
This influences the propagation of plasma and electromagnetic waves
\cite{Movvlprop,Movvlscat,Movcrvl}.
If in the case of a dusty plasma we fix the grain charge, then the grain becomes
entirely defined in terms of its mass and charge, and as such acts just like
another ion in the plasma (albeit of negative charge).
The state of charge neutrality may be written as:
\begin{equation}
-en_{e}+eZ_{1}n_{1}+eZ_{2}n_{2}=0. \label{Movequil}
\end{equation}
Here $n_{e,1,2}$ is the number density of plasma electrons (with
the charge $-e$), and the two ion species (of signed charge $Z_{1}$ and $Z_{2}$), respectively.
For laboratory dusty plasmas, the grain charge is negative (i.e., $Z_2<0$) and large
($|Z_2|\sim 10^2-10^3$), so that an appreciable proportion of the
negative charge in the plasma may reside on the dust particles.
For astrophysical dusty plasmas, $|Z_2|$ may only be of the order of unity,
which is often the case for a canonical bi-ion plasma. In environments
such as the interstellar medium, where the dust grain is in an electron-proton plasma with little ultra-violet 
radiation present, the
dust grains acquire a negative charge. On the other hand, exposure to ultraviolet light from nearby stars
can cause ionization of the grains, leaving a residual positive charge.

In this paper, we investigate the propagation of plane
hydromagnetic waves (Alfv\'en waves and magnetoacoustic waves modified by the presence of a secondary ion or dust 
species in a bi-ion plasma), propagating parallel to the pumped
magnetic field of a large amplitude magnetoacoustic wave. We
generalize Ref.\,\cite{MovHertz} where the second heavy space was
assumed immobile, to the case where both ions are fully mobile. A
further generalization is the inclusion of pressure. The
background magnetic field and plasma density are taken to be
uniform, at frequencies well below the electron plasma and
cyclotron frequencies. First, we find the dispersion equation of
the pump wave, and then concentrate on pump waves of large
wavelength. We then obtain a coupled pair of equations of motion
governing the perturbed plasma. Resonant interactions are sought, and growth rates of
the waves parametrically excited by the pump are calculated and discussed.

\section{Multifluid Model}\label{MovWaveqnsP}
The most general set of equations used to describe two mobile
ion-like species, plus electrons, is a three-fluid model. In this
picture we employ three momentum equations for the electrons and
the two ion species, where we include both the ions' inertia
terms, while ignoring the electron inertia. This is valid if the
frequencies of interest are well below the electron cyclotron frequency.
In addition we use Maxwell's equations and two mass continuity
equations for each of the ion species, ignoring Maxwell's
displacement current.

The primary species (assumed positively charged) shall be denoted with a ``1"
subscript, and the secondary species (either positively or negatively charged) shall be
denoted with a ``2" subscript. In this notation we may write the
current density as
\begin{equation}
{\bf J}= Z_{1}e n_{1}{\bf v}_{1}+ Z_{2}e n_{2} {\bf v}_{2}- e
n_{e}{\bf v}_{e},
\end{equation}
where the two ion species and electrons have velocities denoted by
${\bf v}_{1}$, ${\bf v}_{2}$, and ${\bf v}_{e}$, respectively.
We employ the parameters $\delta_{1}
=n_{e}/Z_{1}n_{1}$ and $\delta_{2}=n_{e}/Z_{2}n_{2}$, which
measure the distribution of charge in the plasma amongst the ions. Employing the
total charge neutrality condition, given in Eq.\,(\ref{Movequil}), we
may write this as
\begin{equation}
\frac{1}{\delta_{1}}+\frac{1}{\delta_{2}}=1,
\end{equation}
in terms of these parameters. Note that in the limit of a
single (primary) ion species, we have $\delta_{1}\to 1$ and
$|\delta_{2}|\to\infty$.

Ignoring collisions, but including the effects of pressure, the
starting equations for the velocities, electric and magnetic
fields and each number density are the momentum equations, the two ion continuity equations, and Ampere's law 
neglecting the displacement current:
\begin{eqnarray}  m_1n_1\frac{d{\bf v}_1}{dt}\al=\al
-\nabla\!p_{1}+Z_{1}en_1\left({\bf E}+{\bf v}_1\times{\bf B}\right)
\label{Movi},  \\
m_2n_2\frac{d{\bf v}_2}{dt}\al=\al-\nabla\!p_{2}+ Z_{2}en_2
\left({\bf E}+{\bf v}_2\times{\bf B}\right)  \label{Movd},  \\
{\bf 0}\al=\al-\nabla\!p_{e}-en_e\left({\bf E}+{\bf v}_e\times{\bf
B}\right) \label{Move},  \\
\frac{\partial n_{1}}{\partial t}\al=\al-{\bf\nabla}\cdot(n_{1}
{\bf v}_{1})
\label{Movcon}, \\
\frac{\partial n_{2}}{\partial t}\al=\al-{\bf\nabla}\cdot(n_{2}
{\bf v}_{2}),
\label{Movcond} \\
{\bf\nabla} \times {\bf B}\al=\al\mu_0 e(Z_{1}n_{1}{\bf v}_1
+Z_{2}n_{2}{\bf v}_{2} -n_{e}{\bf v}_e ). \label{Movmax2}
\end{eqnarray}
Here, $p_{1,2,e}$ are the two heavy species and electron
pressures, and $m_{1,2}$ are the heavy masses. The
magnetic field ${\bf B}$ includes the background magnetic field
${\bf B}_{0}$.

Finally, by assuming either an isothermal or an adiabatic equation
of state, we have
\begin{eqnarray}
\nabla\!p_{\alpha}=U_{\alpha}^2 m_{\alpha}\nabla\!n_{\alpha}
\end{eqnarray}
for each species, where $U_{\alpha}$ are the individual sound
speeds. Though it is not imperative, we shall assume isothermal
changes, which permits us to write down $U_{\alpha}$ in terms of
the plasma temperatures, i.e.,
$U_{\alpha}^{2}=k_{B}T_{\alpha}/m_{\alpha}$
 is the square of each thermal speed, where $T_{\alpha}$ is
each temperature and $k_{B}$ is Boltzmann's constant.

At this point we choose to eliminate the electron variables from
Eqs.\,(\ref{Movi})--(\ref{Movmax2}), and use Faraday's law to eliminate
the electric field ${\bf E}$. There are then two choices for the
way to proceed. We can add and subtract the momentum equations and
deal with a total fluid velocity and the current density ${\bf J}$. Instead
we shall employ the most direct method, which is to deal with the ion
velocities ${\bf v}_{1}$ and ${\bf v}_{2}$ separately, since our
equations will exhibit the most symmetry this way. By employing
the charge neutrality condition to all orders, the following
system of nonlinear partial differential equations is found:
\begin{eqnarray}
\frac{\partial\rho_{1}}
{\partial t}\al=\al-\nabla\cdot(\rho_{1}{\bf v}_{1}),
\label{Movcont}\\
\frac{\partial\rho_{2}}{\partial t}\al=
\al-\nabla\cdot(\rho_{2}{\bf v}_{2}),
\label{Movcontd}\\
\frac{\partial{\bf B}}{\partial t}\al=\al\nabla\!\times\left(
{\bf v}_{1}\times{\bf B}\right)-\frac{B_{0}}{\Omega_{1}}\nabla\!\times
\frac{d{\bf v}_{1}}{d t}\nonumber\\
\al=\al\nabla\!\times\left(
{\bf v}_{2}\times{\bf B}\right)-\frac{B_{0}}{\Omega_{2}}\nabla\!\times
\frac{d{\bf v}_{2}}{d t},\,\,\,\,\,\,\,\,\,\,\,
\label{Movmag1}\\
\delta_{1}\rho_{1}\frac{d{\bf v}_{1}}{dt}\al=\al-
(\delta_{1}U_{1}^{2}+\alpha_{1}^{2})
\nabla\!\rho_{1}-\alpha_{2}^{2}\nabla\!\rho_{2} \nonumber\\
\al+\al \frac{\rho_{2}\Omega_{2}}{B_{0}}({\bf v}_{1}-{\bf v}_{2})
\times{\bf B}+\frac{1}{\mu_{0}}(\nabla\!\times{\bf B})
\times{\bf B},\label{Moveul}\\
\delta_{2}\rho_{2}\frac{d{\bf v}_{2}}{dt}\al=\al-
(\delta_{2}U_{2}^{2}+\alpha_{2}^{2})\nabla\!\rho_{2}
-\alpha_{1}^{2}\nabla\!\rho_{1} \nonumber\\
\al+\al\frac{\rho_{1}\Omega_{1}}{B_{0}}({\bf v}_{2}-{\bf v}_{1})
\times{\bf B}+\frac{1}{\mu_{0}}(\nabla\!\times{\bf B}) \times{\bf
B}.\,\,\,\,\,\,\,\,\,\, \label{Moveuld}
\end{eqnarray}
In Eqs.\,(\ref{Movcont})--(\ref{Moveuld}) $\rho_{1,2}=m_{1,2}n_{1,2}$
are the densities of each massive component of the plasma and
$\Omega_{1,2}=Z_{1,2} eB_0/m_{1,2}$ are the corresponding (signed)
cyclotron frequencies, with $B_{0}=|{\bf B}_{0}|$. Also
$\alpha_{1}^{2}=Z_{1}U_{e}^{2}m_{e}/m_{1}$ and
$\alpha_{2}^{2}=Z_{2}U_{e}^{2}m_{e}/m_{2}$ are pseudo-squared thermal
speeds associated with the electron pressure (if $Z_{2}<0$ we
allow $\alpha_{2}$ to be imaginary). The presence of both $\nabla\!\rho_{1}$
and $\nabla\!\rho_{2}$ in both equations of motion, is a consequence
of the elimination of the electron variables from the equations. To be more explicit,
the electron density fluctuations generate fluctuations in both the ion densities
through the charge neutrality condition, which produces this coupling in the
momentum equations.
The second term on the right hand side of Eq.\,(\ref{Movmag1}) is the Hall term; it
is important when the wave frequency is comparable to
either of the cyclotron frequencies. Also, we see that the two
species are strongly coupled through the momentum equations, via
the third term on the right hand side in each. This is an
additional Hall-type term associated with the relative motion of
the two species. Note Eqs.\,(\ref{Movcont})--(\ref{Moveul}) reduce, in
the case of a single species plasma ($\delta_{1}\to 1,\,\,
|\delta_{2}|\to\infty$, assuming local charge neutrality is
maintained), to the equations used in Ref.\,\cite{Movmio,Movmjolh76,Movspsh,Movmjwy86,Movovendon}
where nonlinear Alfv\'en waves were investigated. They are then known as the (collisionless)
Hall-MHD (magnetohydrodynamics) equations.

The neglect of the displacement current in Maxwell's equation is justified when the electron current is much greater
than the displacement current, which leads to the conditions
\begin{equation}
\delta_{1}\gg\frac{\omega}{\Omega_{1}}\frac{v_{A1}^2}{c^2},\,\,\,
|\delta_{2}|\gg\frac{\omega}{|\Omega_{2}|}\frac{v_{A2}^2}{c^2},\,\,\,
\label{MovCond1}\end{equation}
where $\omega$ is the wave frequency and $v_{A1}, v_{A2}$ are the Alfv\'en speeds
associated with each ion.
The charge neutrality condition implies that
the wave frequencies are restricted to the regime $\omega\ll\omega_{pe}$,
which may be restated as
\begin{equation}
|\delta_{1,2}| \gg \frac{|\Omega_{1,2}|}{\Omega_{e}}
\left(\frac{\omega}{\Omega_{1,2}}\right)^2\frac{v_{A1,2}^2}{c^2}.
\label{MovCond2}\end{equation}
There exists a wide range of physical environments for which all of the
above mentioned conditions are met.

\section{Magnetoacoustic Pump Wave}\label{Movparapump}
Suppose there is a constant background magnetic field in
the $z$-direction, given by $B_{0}\hat{{\bf z}}$.
We now periodically pump the field, with a
periodic modulation:
\begin{equation}
B_{0}[1+\bar{\varepsilon}b\cos(\omega_{0}t)]\hat{{\bf z}},
\label{MovinitB}\end{equation}
where $\omega_{0}$ is the pump frequency.
Here $\bar{\varepsilon}$ is a constant dimensionless quantity
which determines the amplitude of the pump. More precisely, $\bar{\varepsilon}$
is an expansion parameter, which permits us to keep track of our terms,
by matching powers of $\bar{\varepsilon}$.
The parameter $b$ on the other hand is a dimensionless quantity that we include to capture
any necessary frequency information, i.e., $b=b(\omega_{0})$,
such that the average of $b$ over $\omega_{0}$ is $O(1)$.
By specifying a particular choice of normalization condition, we may solve for $b$.
We later in fact impose the condition that the energy density in the pump system is a constant over $\omega_0$, and 
calculate the resulting $b$.
Since we are interested in a large amplitude pump wave, $\bar{\varepsilon}$ will
typically be $O(10^{-2}-10^{-1})$.

However, since this field has no spatial dependence,
it does not satisfy our wave equations.
Hence this is an approximation to a wave with a large
wavelength. We must therefore modify the pump with an envelope of
some wavelength. Since the pump wave magnetic field points
in the $z$-direction, the wavevector should point perpendicular
to this to satisfy $\nabla\cdot{\bf B}=0$.
For a planar geometry plasma, we choose the axes such that the wave
varies in the $x$-direction, with wavenumber $k_{0}$.
By denoting the pump magnetic field (and all subsequent
pump fields) with a ``0" superscript, we have
\begin{equation}
{\bf B}^{(0)}=B_{0}[1+\bar{\varepsilon}b\cos(k_{0}x)
\cos(\omega_{0}t)]\hat{\bf z}.
\label{Movpump}\end{equation}
In a cylindrical plasma, with $r$ denoting the
radial distance from the cylindrical axis, we have
\begin{equation}
{\bf B}^{(0)}=B_{0}[1+\bar{\varepsilon}bJ_{0}(k_{0}r)
\cos(\omega_{0}t)]\hat{\bf z},
\label{MovpumpJ}
\end{equation}
 where $J_{0}$ is the 0th
order Bessel function of the first kind. Note that
$b=b(k_{0},\omega_{0})$ in Eqs.\,(\ref{Movpump}) and (\ref{MovpumpJ}).

In the absence of a wave, we shall suppose
that the plasma is stationary, with ion densities $\rho_{01,2}$
and charge ratios of $\delta_{01,2}$.
The effect of pumping will be to modify the velocity, density
and charge imbalance to order $\bar{\varepsilon}$, so that
\begin{eqnarray}
{\bf v}_{1}^{(0)}\al = \al\bar{\varepsilon}
{\bar{\bf v}}_{1} , \,\,\,\,\,\,\,
\,\,\,\,\,\,\,\,\,\,\,\,
{\bf v}_{2}^{(0)}=\bar{\varepsilon}{\bar{\bf v}}_{2},\\
\rho^{(0)}_{1}\al = \al\rho_{01}+\bar{\varepsilon}\bar{\rho}_{1}
\label{Movrho} , \,\,\,\,\,\,
\rho^{(0)}_{2}=\rho_{02}+\bar{\varepsilon}\bar{\rho}_{2},
\label{Movrhod}\\
\delta^{(0)}_{1}\al =\al \delta_{01}+\bar{\varepsilon}
\bar{\delta}_{1} , \,\,\,\,\,\,\,
\delta^{(0)}_{2}=\delta_{02}+\bar{\varepsilon}\bar{\delta}_{2}.
\end{eqnarray}
These quantities define the pump wave, and we now proceed to
solve the equations (\ref{Movcont})--(\ref{Moveuld}) to order $\bar{\varepsilon}$
(i.e., the linear solution).

By inspecting the form of the magnetic field in Eq.\,(\ref{Movpump}) [or (\ref{MovpumpJ})],
and Eqs.\,(\ref{Moveul}) and (\ref{Moveuld}),
we note that the velocity cannot have a $z$-component. Hence, the velocities of the ion species must be of the form
\begin{equation}\bar{\bf v}_{1}=(\bar{v}_{x1},\bar{v}_{y1},0)
\,\, , \,\,\,\,\,
\bar{\bf v}_{2}=(\bar{v}_{x2},\bar{v}_{y2},0)
\label{Movtransvel}\end{equation}
for planar waves.
From the resulting pair of equations we ascertain the form of each
velocity component and density:
\begin{eqnarray}
\bar{v}_{x1,2}\al=\al A_{x1,2}\sin(k_{0}x)\sin(\omega_{0}t)\label{Movv_x}, \\
\bar{v}_{y1,2}\al=\al A_{y1,2}\sin(k_{0}x)\cos(\omega_{0}t)\label{Movv_y}, \\
\bar{\rho}_{1,2}\al=\al R_{1,2}\cos(k_{0}x)\cos(\omega_{0}t)\label{Movrho1},
\end{eqnarray}
where $A_{x1,2}, A_{y1,2}$ and $R_{1,2}$ are amplitudes independent
of space
and time, to be determined.
Then, from the continuity equations, we have
\begin{equation} R_{1}=\rho_{01}\frac{k_{0}}{\omega_{0}}A_{x1}
, \,\,\,\,
R_{2}=\rho_{02}\frac{k_{0}}{\omega_{0}}A_{x2}.
\label{Movconamp}\end{equation}

For the cylindrical plasma case, we make the substitutions:
$x\to r$, $y\to\phi$, $\cos\to J_{0}$ and $\sin\to J_{1}$,
where $\phi$ is the azimuthal angle and $J_{1}$
is the 1st order Bessel function of the first kind.
Then it is found that all the relationships
(\ref{Movtransvel})--(\ref{Movconamp}) still hold.
It follows that the modes in the planar and cylindrical plasmas
satisfy the same dispersion equation.

\begin{figure}[t]
\centerline{\includegraphics[width=10cm]{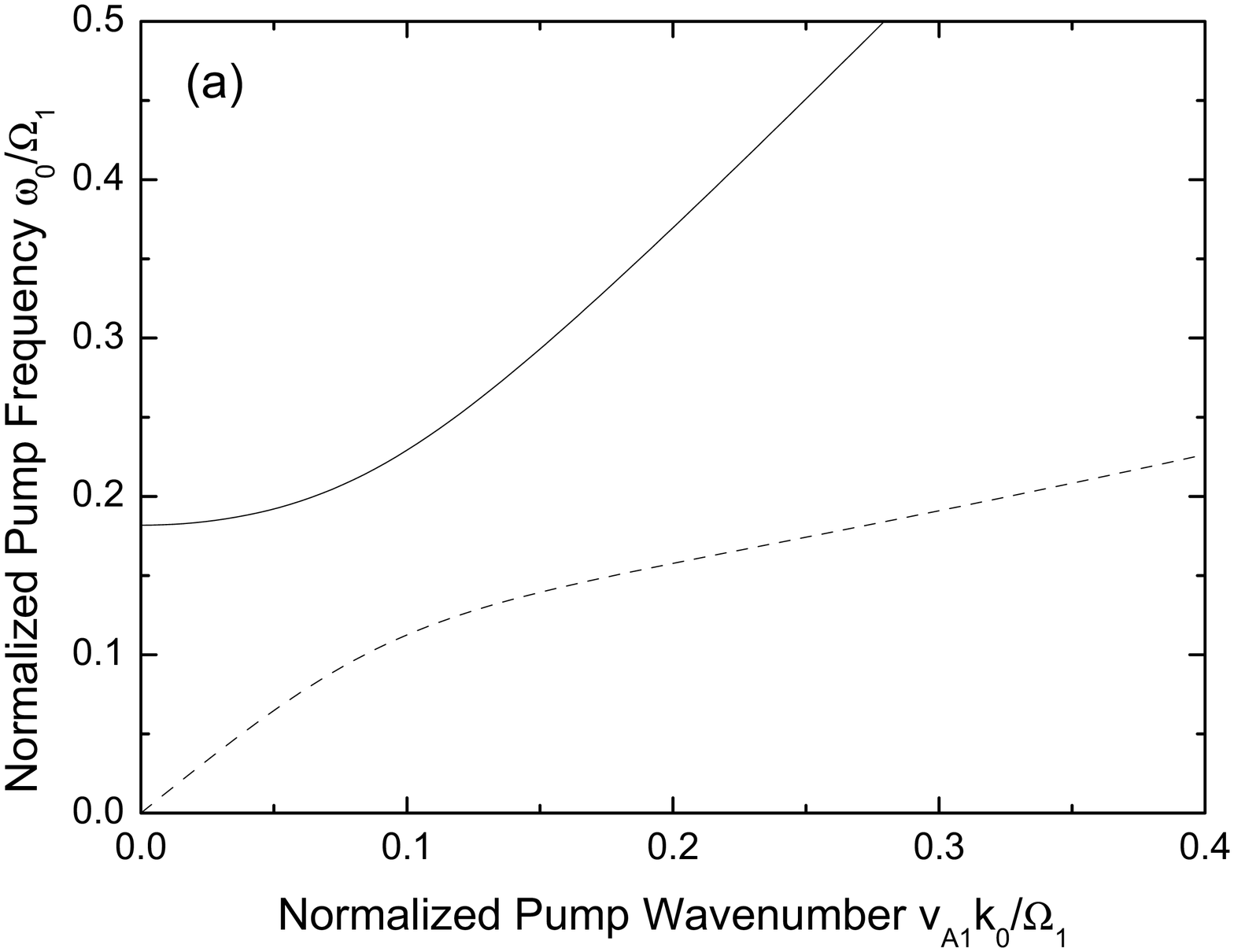}}
\centerline{\includegraphics[width=10cm]{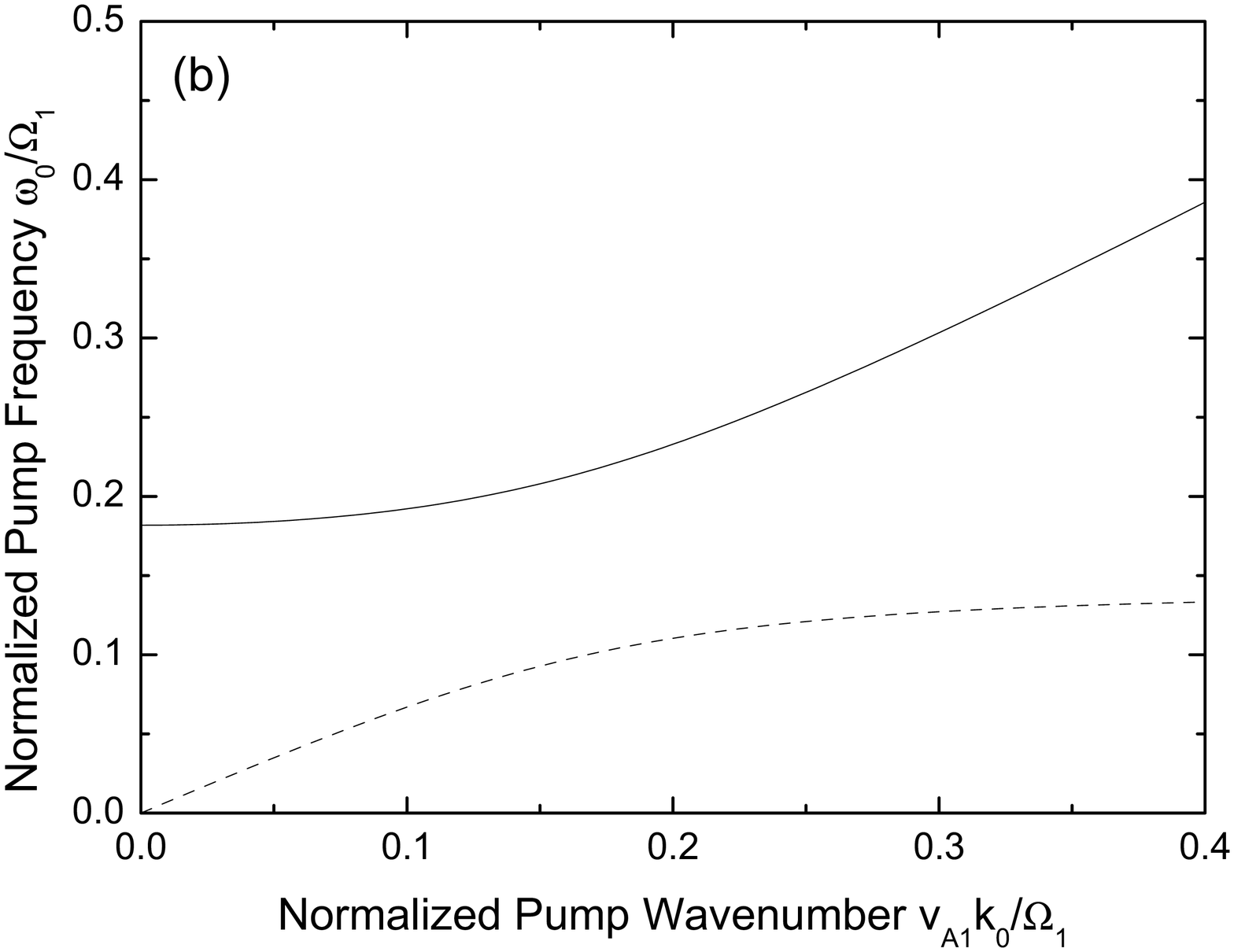}}
\caption{
The normalized pump frequency $\omega_{0}/\Omega_{1}$
versus normalized pump wavenumber $v_{A1}k_{0}/\Omega_{1}$ with
$\delta_{01}=1.1$ and $\Omega_{2}/\Omega_{1}=0.1$.
(a) The warm plasma, with
$\beta_{1}=0.8,\, B_{1}=1.5$ and $U_2/U_1=1/2$.
(b) The cold plasma, with
$\beta_{1}=B_{1}=U_2=0$.}\label{Mov1}
\end{figure}

Upon substitution of Eqs.\,(\ref{Movv_x})--(\ref{Movconamp}) into Eqs.\,(\ref{Movcont})--(\ref{Moveuld}),
written to first order in 
$\bar{\varepsilon}$, the following dispersion equation of the pump wave is obtained:
\begin{equation}
\omega_{0}^{2}\,(\omega_{0}^{2}-\omega_{c}^{2})
=W\,k_{0}^{2}\,(\omega_{0}^{2}-X-Y k_{0}^{2}).
\end{equation}
Here, we have defined
\begin{eqnarray}
\omega_{c}\al=\al\Omega_{1}/\delta_{02}
+\Omega_{2}/\delta_{01}, \label{Movomega_c}\\
W\al=\al v_{A1}^{2}(1+\delta_{01}^{2}\beta_{1}+\delta_{01}B_{1})
/\delta_{01}^{2}\nonumber\\
\al+\al v_{A2}^{2}(1+\delta_{02}^{2}\beta_{2}+\delta_{02}B_{2})
/\delta_{02}^{2},\\
X\al=\al\omega_{c}(
v_{A1}^{2}\Omega_{2}(1+\beta_{1}+\delta_{01}B_{1})/\delta_{01}\nonumber\\
\al+\al v_{A2}^{2}\Omega_{1}\beta_{2}/\delta_{02})/W, \\
Y\al=\al v_{A1}^{2}v_{A2}^{2}(\beta_{2}
(1+\delta_{01}B_{1})/\delta_{01}^{2}\nonumber\\
\al+\al \beta_{1}(1+\delta_{02}B_{2})/\delta_{02}^{2}
+\beta_{1}\beta_{2})/W,
\label{Movcutandres}\end{eqnarray}
where $|\omega_{c}|$ is a hybrid cutoff frequency as
$k_{0}\to 0$.
Also we have introduced $\beta_{1}:=U_{1}^{2}/v_{A1}^{2}$,
 $\beta_{2}:=U_{2}^{2}/v_{A2}^{2}$,
 $B_{1}:=\alpha_{1}^{2}/v_{A1}^{2}$ and
 $B_{2}:=\alpha_{2}^{2}/v_{A2}^{2}$
which allow us to specify the cold/warm plasma regimes,
with $B_{1}$ and $B_{2}$ related through:
$\delta_{0 1}B_{1}=\delta_{0 2}B_{2}$.
Note that as $\delta_{01}\to 1\, , \,\,|\delta_{02}|\to\infty$
(i.e., a single species)
we are left with the nondispersive relation
\begin{equation}
\omega_{0}=\sqrt{c_{s}^{2}+v_{A1}^{2}}\,k_{0},
\end{equation}
where $c_{s}^{2}=U_{1}^{2}+\alpha_{1}^{2}$ is the combined sound speed
in this limit.
This is the familiar fast magnetoacoustic wave characteristic for
${\bf k}\perp {\bf B}_{0}$. However, in the presence of a secondary ion species
the fast magnetoacoustic wave
gains an additional mode and the relationship is dispersive,
see Fig.\,\ref{Mov1}.

It is easy to show that the parameters $X$ and $W$ are always positive,
while $Y=0$ for the cold plasma and is positive
if any of $\beta_{1}, \beta_{2}, B_{1}, B_{2}$ are nonzero.
Although Fig.\,\ref{Mov1} is for a plasma in which
the secondary ion species is positive, the same basic
qualitative features are present when the second ion species
is negative. For the warm plasma the effect
of the $Y k_{0}^{2}$ term in the dispersion relation
is to cause the lower branch to increase without bound
as $k_{0}\to\infty$, as indicated in Fig.\,\ref{Mov1}(a).
Physically, it can be thought of as the fast magnetoacoustic
mode being converted into an acoustic mode for large $k_{0}$
due to the inclusion of pressure.
However, if the plasma is cold (i.e., $Y=0$) then the lower
branch experiences a resonance as $k_{0}\to\infty$, as
indicated in Fig.\,\ref{Mov1}(b). In this limit
$X$ becomes the square of a resonance frequency, given by
\begin{equation}
X\to\omega_{r}^{2}=\frac{\omega_{c}v_{A1}^{2}\Omega_{2}\delta_{02}}
{\delta_{02}v_{A1}^{2}/\delta_{01}
+\delta_{01}v_{A2}^{2}/\delta_{02}},
\end{equation}
which gives rise to a forbidden frequency region between
$|\omega_{r}|$ and $|\omega_{c}|$, since  $|\omega_{r}|<|\omega_{c}|$
whenever $\delta_{01}\neq 1$.

The ratio of the
$x$-velocity amplitudes $A_{x1,2}$ to the magnetic field parameter $b$
 are of particular importance
to our later discussion, so we briefly discuss them here.
In terms of the frequency and wavenumber, we find,
\begin{equation}
\frac{A_{x1}}{b}=\frac{\omega_{0}}{k_{0}A_{1}},\,\,\,\,
\frac{A_{x2}}{b}=\frac{\omega_{0}}{k_{0}A_{2}},
\label{Movxvel1}\end{equation}
where
\begin{eqnarray}
A_{1}\al :=\al\frac{1}{\delta_{01}}+\frac{1}{\delta_{02}}
\frac{\Omega_{2}}{\Omega_{1}}\frac{U_{1}^{2}k_{0}^{2}+\Omega_{1}
\omega_{c}-\omega_{0}^{2}}
{U_{2}^{2}k_{0}^{2}+\Omega_{2}\omega_{c}-\omega_{0}^{2}},\nonumber\\
A_{2}\al:=\al\frac{1}{\delta_{02}}+\frac{1}{\delta_{01}}
\frac{\Omega_{1}}{\Omega_{2}}\frac{U_{2}^{2}k_{0}^{2}+\Omega_{2}
\omega_{c}-\omega_{0}^{2}}
{U_{1}^{2}k_{0}^{2}+\Omega_{1}\omega_{c}-\omega_{0}^{2}}.
\label{Movxvel2}\end{eqnarray}
Finally, to complete the set of pump quantities, we find that
the velocity in the $y$-direction is given by
\begin{equation}
\frac{A_{y 1}}{A_{x 1}}
=\frac{(U_{2}^{2}k_{0}^{2}-\omega_{0}^{2})\Omega_{1}
-(U_{1}^{2}k_{0}^{2}-\omega_{0}^{2})\Omega_{2}}
{\delta_{02}\omega_{0}(
U_{2}^{2}k_{0}^{2}+\Omega_{2}\omega_{c}-\omega_{0}^{2})}=F_1,
\label{MovF1}\end{equation}
with $A_{y 2}$ given similarly.
This is quite different to the single ion species case, where
$A_{y}\equiv 0$.

\begin{figure}[h]
\centerline{\includegraphics[width=10cm]{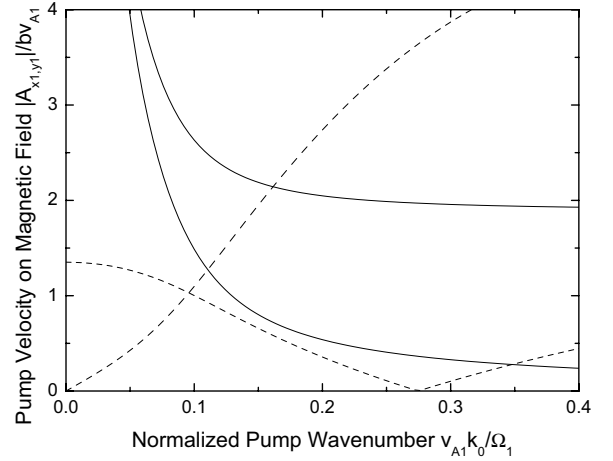}}
  \caption{The ratio of normalized $x$-velocity
amplitude and $y$-velocity amplitude
to magnetic field $A_{x1,y1}/b v_{A1}$
versus normalized pump wavenumber $v_{A1} k_0/\Omega_{1}$,
with $\delta_{01}=1.1,\, \Omega_{2}/\Omega_{1}=0.1,\, \beta_1=0.8,\,
B_1=1.5$ and $U_2/U_1=1/2$. The solid curve is for the upper branch
of the dispersion relation, the dashed curve is for the lower
branch of the dispersion relation. Note that on the vertical axis we plot
the absolute value of this ratio, since we are interested here in the ratio of {\em amplitudes}.
Consequently one of the two dashed curves experiences a discontinuity in its first derivative when
it touches the horizontal axis.}\label{Mov2}
\end{figure}

The behavior of $|A_{x,y}|/b$ is dramatically altered by
the choice of branch from the dispersion relation; see
Fig.\,\ref{Mov2}, where for brevity we plot $A_{x1,y1}$ only.
We see that as $k_0\to\infty$, the $x$-component approaches a finite positive value
while the $y$-component approaches zero for the upper branch of the dispersion relation.
On the other hand the $x$-component approaches large values and the $y$-component passes through zero
for the lower branch of the dispersion relation.

\begin{figure}[h]
\centerline{\includegraphics[width=10cm]{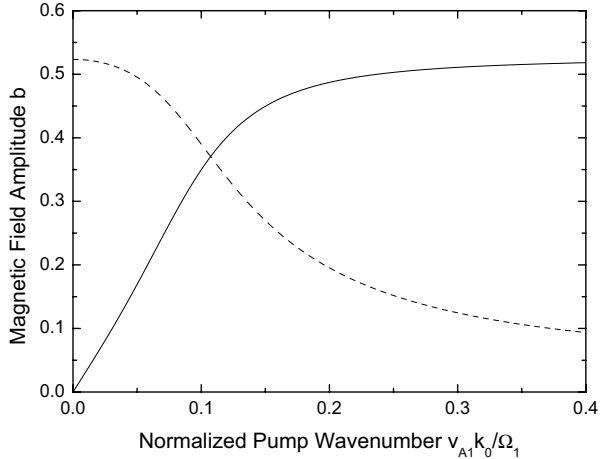}}
  \caption{The magnetic field amplitude $b$
versus normalized pump wavenumber $v_{A1} k_0/\Omega_{1}$
due to the normalization condition (\ref{Movbnorm}),
with $\delta_{01}=1.1,\, \Omega_{2}/\Omega_{1}=0.1,\, \beta_1=0.8,\,
B_1=1.5$ and $U_2/U_1=1/2$. The solid curve is for the upper branch
of the dispersion relation, the dashed curve is for the lower
branch of the dispersion relation.}\label{Mov3}
\end{figure}

Note that the two branches stemming from the upper branch of the dispersion
relation become singular as $k_0\to 0$. This motivates the need for a
normalization condition relating the pump field amplitudes of the velocities
$A_{x1,2}, A_{y1,2}$, magnetic field $b$, and
density $R_{1,2}$. A reasonable
way to proceed is to assume that as we vary $\omega_{0}$
the energy density $E_{D}$ in the pump system is fixed. The appropriate relation in a fluid description for general 
magnetoacoustic waves is (e.g.\ \cite{Movmelrose}),
\begin{eqnarray}
E_{D}\al=\al\frac{\bar{B}^{2}}{2\mu_{0}}+\frac{\epsilon_{0}\bar{E}^{2}}{2}
+\frac{1}{2}\rho_{01}\left(A_{x1}^{2}+A_{y1}^{2}\right)\nonumber\\
\al+\al\frac{1}{2}\rho_{02}\left(A_{x2}^{2}+A_{y2}^{2}\right)
+\frac{1}{2}\sum_{\alpha}
\frac{U_{\alpha}^{2}}{\rho_{0\alpha}}R_{\alpha}^{2},
\end{eqnarray}
where $\bar{B}=B_{0}b$, and
the summation in the final term is over each plasma species.
In our approximations the electric energy term may be neglected
compared to the magnetic energy.
By switching to dimensionless units and setting
$\mu_{0}E_{D}/B_{0}^{2}\equiv 1$, then all our variables
are given uniquely in terms of the basic plasma parameters.
By using the expressions for $A_{x1,y1}$ and $A_{x2,y2}$ we may solve for $b$.
The result is
\begin{equation}
b=\frac{\sqrt{2}}
{\sqrt{1+\frac{\omega_0^2(1+F_1^2)}{k_0^2 v_{A1}^2 A_1^2}
 \frac{\omega_0^2(1+F_2^2)}{k_0^2 v_{A2}^2 A_2^2}+\frac{\beta_1}{A_1^2}
 +\frac{\beta_2}{A_2^2}+B_1\delta_{01}}}
\label{Movbnorm} \end{equation}
[using $F_{1,2}$ from Eq.\,(\ref{MovF1})].
A plot of $b$ versus wavenumber is given in Fig.\,\ref{Mov3} for each branch $\omega_{0}(k_0)$.
It is important to note that $b=O(v_{A1}k_{0}/\Omega_1)$ as $k_0\to 0$ for the upper branch.
It follows that all the velocity amplitudes are well behaved near $k_{0}\simeq 0$.

The remaining pump quantities to solve for are the charge ratios $\delta^{(0)}_{1,2}$.
These may be obtained by imposing the charge neutrality condition
on both the background plasma and on the perturbations to the equilibrium.
This is justified since we are interested only with frequencies that are much less than that
of the electron plasma frequency. This means the electrons
have sufficient time to respond to the perturbations from the equilibrium
and neutralize the plasma at each point in space.
Thus, on the perturbed plasma we impose the condition,
 \begin{equation}
-e\bar{n}_{e}+Z_{1} e \bar{n}_{1}+Z_{2}e \bar{n}_{2}=0,
\end{equation}
where we have ignored the effects of dust (or ion) charging, that is, $Z_{2}$ is constant.
It is then straightforward to obtain the charge ratios in terms of the densities,
and subsequently in terms of the velocity amplitudes and $b$, as follows:
\begin{eqnarray}
\delta^{(0)}_{1}=\delta_{01}\left(1+\bar{\varepsilon}(b-\frac{k_{0}}
{\omega_{0}}A_{x1})\cos(k_{0}x)\cos(\omega_{0}t)\right),\nonumber\\
\delta^{(0)}_{2}=\delta_{02}\left(1+\bar{\varepsilon}(b-\frac{k_{0}}
{\omega_{0}}A_{x2})\cos(k_{0}x)\cos(\omega_{0}t)\right).
\label{Movcharge2}\end{eqnarray}
This implies that the amount of charge that resides on each species
is temporally coupled to the other.

Using the derived set of quantities characterizing the pump wave we may, in principle,
investigate the possibility of excited modes propagating at arbitrary angles with
respect to the $z$-axis, as in Ref.\,\cite{MovCramer76}. However, by removing the $x$-dependence
in the pump waves our calculations become much more tractable, with
the requirement that our excited waves are restricted to {\it parallel} propagation
along the $z$-axis, see Ref.\,\cite{MovCramer}.

For an infinitely extended medium, the perpendicular wavenumber of the excited waves is then zero. However, this would 
clearly be inconsistent with wavenumber matching of a propagating pump with finite perpendicular wavenumber. We 
therefore envisage this analysis to apply to a standing wave (in the transverse direction) pump in a finite geometry 
system, such as a cylindrical or toroidal laboratory plasma, or an astrophysical magnetic flux tube; by seeking 
localized solutions for distances transverse to the magnetic field
such that $k_{0}x=O(\bar{\varepsilon})$ in a planar plasma,
or $k_{0}r=O(\bar{\varepsilon})$ in a cylindrical plasma,
we can recover the initially proposed form for the magnetic field
Eq.\,(\ref{MovinitB}). In other words, we are taking $x$ small rather than $k_0$ small, to ensure that the fields
will look approximately uniform in the perpendicular direction, i.e.\ that the wavelength is effectively very long on 
the length scale of interest. In the parallel direction, however, the pump fields are uniform, we place no such 
restriction on distances and excited wavelengths, and we can apply wavenumber matching of the excited wave fields in 
that direction, as discussed in Section VI.

Then in the region near the $z$-axis we have
\begin{eqnarray}
\left[{\bf B}^{(0)}\right]_{k_{0}x\sim\bar{\varepsilon}}\al
=\al B_{0} (1+\bar{\varepsilon}b\cos(\omega_{0}t))
\hat{\bf z},\label{Movmag2}\\
\left[{\bf v}_{1}^{(0)}\right]_{k_{0}x\sim\var}\al=\al{\bf 0},
\,\,\,\,\,\,\,\,\,
\left[{\bf v}_{2}^{(0)}\right]_{k_{0}x\sim\var}={\bf 0},\\
\left[\rho^{(0)}_{1,2}\right]_{k_{0}x\sim\var}\al=\al\rho_{01,2}
(1+\bar{\varepsilon}\frac{k_{0}}{\omega_{0}}
A_{x1,2}\cos(\omega_{0}t))\label{Movrho2},\\
\left[\delta^{(0)}_{1,2}\right]_{k_{0}x\sim\bar{\varepsilon}}\al
=\al\delta_{01,2}(1+\bar{\varepsilon}(b-\frac{k_{0}}{\omega_{0}}A_{x1,2})\cos(\omega_0t)),\,\,\,\,\,\,\,
\end{eqnarray}
correct to $O(\var)$.
Since the magnetic field has now simply a
periodic variation in time it resembles a canonical parametric pump. The problem then resembles a parametric 
amplifier, such as a harmonic oscillator with a time-varying spring constant.
For use in the excited wave equations we must also compute the derivatives near the $z$-axis, given here:
\begin{eqnarray}
\left[\frac{\partial{\bf B}^{(0)}}{\partial x}
\right]_{k_{0}x\sim\bar{\varepsilon}}\al=\al {\bf 0},\\
\left[\frac{\partial{\bf v}^{(0)}_{1}}{\partial x}
\right]_{k_{0}x\sim\bar{\varepsilon}}\al=\al
\bar{\varepsilon}k_{0}(A_{x1}\sin(\omega_{0}t),
A_{y1}\cos(\omega_{0}t),0),\label{Movvelderiv}\\
\left[\frac{\partial{\bf v}^{(0)}_{2}}{\partial x}
\right]_{k_{0}x\sim\bar{\varepsilon}}\al=\al
\bar{\varepsilon}k_{0}(A_{x2}\sin(\omega_{0}t),
A_{y2}\cos(\omega_{0}t),0),\,\,\,\,\,\,\,\,\label{Movvelderiv2}\\
\left[\frac{\partial{\rho}^{(0)}_{1,2}}{\partial x}
\right]_{k_{0}x\sim\bar{\varepsilon}}\al=\al
\left[\frac{\partial{\delta}^{(0)}_{1,2}}{\partial x}
\right]_{k_{0}x\sim\bar{\varepsilon}}=0.\label{Movrhoderiv}
\end{eqnarray}
Note that the velocity has a finite derivative, even though the velocity itself is zero in
this regime.

\section{Perturbation Method}\label{MovPertmeth}
Next we wish to test the stability of the self consistent linear solution given
in Eqs.\,(\ref{Movmag2})--(\ref{Movrhoderiv}), now regarded as a finite-amplitude pump wave, to the excitation of waves 
propagating along the magnetic field direction.
We must however be careful since this is only a linearized pump solution.
To proceed we will adopt the basic methodology of Refs.\,\cite{MovVahal,MovCramer},
but shall attempt to refine the argument.

The basic technique is to perturb each quantity, i.e., ${\bf B}^{(0)}$,
${\bf v}^{(0)}$, $\rho^{(0)}$, $\delta^{(0)}$, by an arbitrarily
small amount as measured by an expansion parameter $\varepsilon'$.
For example, take an arbitrary field, say $X$,
with a known linear representation of the pump field:
$X^{(0)}=X_{0}+\bar{\varepsilon}\bar{X}$.
Then Refs.\,\cite{MovVahal,MovCramer} proceed by perturbing this
quantity in the following way:
\begin{equation}
X=X^{(0)}+\varepsilon'X'.
\end{equation}
However, since $X^{(0)}$ has neglected terms of
$O(\bar{\varepsilon}^{2})$,
the term of $O(\varepsilon')$ is even more so negligible in this
expansion
(recall that $\varepsilon'$ is arbitrarily small).
The problem lies in the fact that we are ultimately interested in the
stability of the {\em exact} pump solution, say
$Y^{(0)}=X_{0}+\bar{\varepsilon}\bar{Y}$, with
\begin{equation}
\bar{Y}=\bar{X}+O(\bar{\varepsilon}).
\end{equation}
Thus, we shall perform the following expansion instead:
\begin{equation}
X=Y^{(0)}+\varepsilon'X'.
\end{equation}
Upon substitution into our set of nonlinear differential equations
(\ref{Movcont})--(\ref{Moveuld}), we obtain a set of equations
of the following structure (no approximations):
\begin{eqnarray}
\bar{\varepsilon}A(\bar{Y})+
\var^{2}B(\bar{Y})\al+\al\varepsilon'C(X')+{\varepsilon'}^{2}D(X')\nonumber\\
\al=\al\var\varepsilon'E(\bar{Y},X'),
\label{MovABCD}\end{eqnarray}
where $A, B, C, D, E$ are differential operators acting on their respective
arguments which we may treat as functions
(for brevity we have suppressed their dependence on fields other than $X$).
Now $\var A(\bar{Y})+\var^{2}B(\bar{Y})=0$
by definition of an exact pump solution.
Also, the term of $O({\varepsilon'}^{2})$ can certainly be neglected.
Next, we Taylor expand $E(\bar{Y},X')$ and write
\begin{equation}
E(\bar{Y},X')= E(\bar{X},X')+O(\bar{\varepsilon}).
\end{equation}
We are then permitted to neglect this term of $O(\bar{\varepsilon})$
when inserted back into Eq.\,(\ref{MovABCD}).
This gives us a set of equations for the perturbation $X'$,
in terms of the $\bar{X}$, to sufficient accuracy, i.e.,
\begin{equation}
C(X')=\bar{\varepsilon}E(\bar{X},X').
\label{Movpertmeth}\end{equation}
Note that $\bar{X}$ has served two purposes.
First, it gives us an approximate description of the
corresponding exact solution $Y^{(0)}=X_{0}+\var\bar{Y}$, and second,
it allows us to investigate its stability
without ever having to find $\bar{Y}$.

\section{Equations of Motion and Natural Modes}
Our task then is to ascertain the primed variables
which are the excited fields.
As stated previously, we are interested only in plane waves that travel parallel to the $z$-axis. In this case,
any longitudinal components will decouple completely and merely describe a linear
acoustic wave, so we may set the $z$-component of the primed velocity to zero.
Hence, both primed velocities have the following form:
\begin{equation}{\bf v}_{1,2}'=(v_{x1,2}'(z,t),v_{y1,2}'(z,t),0).
\end{equation}
Thus, $\nabla\cdot{\bf v}_{1}'=\nabla\cdot{\bf v}_{2}'=0$,
and from each of the two continuity equations we have
\begin{equation}
\frac{\partial\rho_{1,2}'}{\partial t}=-\var\rho_{1,2}'
\left[\nabla\cdot\bar{\bf v}_{1,2}\right]_{k_0 x\sim\var},
\end{equation}
where we have imposed the $k_{0}x=O(\bar{\varepsilon})$ [or $k_{0}r=O(\bar{\varepsilon})$]
condition. Hence we have $\rho_{1}'=\rho_{2}'=0$, i.e., there are no
perturbations in the densities of the excited fields.
From this it also follows that $\delta_{1}'=\delta_{2}'=0$.

By taking the derived expressions for ${\bf B}$, ${\bf v}$,
${\rho}$ and $\delta$ and substituting into Eqs.\,(\ref{Movmag1}) \& (\ref{Moveul}),
and using the stability analysis procedure conveyed in Eq.\,(\ref{Movpertmeth}), we obtain
\begin{eqnarray}
\al\al\frac{\partial{\bf B}'}{\partial t}-
\nabla\times({\bf v}_{1}'\times{\bf B}_0)
+\frac{B_0}{\Omega_1}\nabla\times\frac{\partial{\bf v}_{1}'} {\partial t}
\nonumber\\ \al=\al\! \bar{\varepsilon}\left[\nabla\times\left({\bf
v}_{1}'\times\bar{{\bf B}} +\bar{\bf v}_{1}\times{\bf
B}'-\frac{B_{0}}{\Omega_{1}}v_{x1}' \frac{\partial\bar{\bf
v}_{1}}{\partial x}\right) \right]_{k_{0}x\sim\bar{\varepsilon}}
\label{Movmag3} \\
\al\rho_{01}\al\delta_{01} \frac{\partial{\bf v}_{1}'}{\partial t}-
\rho_{02}\Omega_{2}({\bf v}_{1}'-{\bf v}_{2}')
\times\hat{\bf z}-\frac{1}{\mu_{0}}
(\nabla\times{\bf B}')\times{\bf B}_{0}\nonumber\\
\al=\al\!\var\left[-\delta_{01}\rho_{01}v_{x1}'\frac{\partial\bar{v}_{1}}{\partial x}
+\Omega_{2}\bar{\rho}_{2}({\bf v}_{1}'-{\bf v}_{2}')\times\hat{\bf z}\right]_{k_{0}x\sim\var}.
\end{eqnarray}
Here the terms of $O(\bar{\varepsilon})$, which occur on the right hand side,
should be thought of as driving terms from the pump wave.
Note that there exist two further equations of motion
under the index interchange $1\leftrightarrow 2$. Note also that there are no acoustic terms on the left hand sides of 
these equations because of the decoupling of the longitudinal motions: however there are still acoustic influences in 
the pump fields on the right hand sides of the equations.

Now in order to treat the $x$ and $y$ components of the ${\bf B}$ and ${\bf v}$ vectors
on an equal footing we form a complex vector out of each component of these partial differential equations,
utilizing the variables $v_{\pm}=v_{x}\pm i v_{y}$. The spatial variation is
assumed to be periodic in the $z$-direction with wavenumber $k$, as follows: $\exp(i k z)$.
Given this, the following pair of linear ordinary differential equations are obtained:
\begin{eqnarray}
\!\al\al\frac{\partial B_{\pm}}{\partial t}- i  k  B_{0}v_{\pm 1}\mp
\frac{k B_{0}}{\Omega_1}\frac{\partial v_{\pm 1}}{\partial t}\nonumber\\
\al=\al i\bar{\varepsilon}\bigg{[}\left(B_{0}b k v_{\pm 1}
+k\frac{B_{0}k_{0}A_{y1}}{\Omega_1}v_{x1}'\pm
k_0 A_{y1}B_{x}'\right)\cos(\omega_0 t)\,\,
\nonumber\\
\al\mp\al A_{x1}\left(k_0 B_{y}'+i
k\frac{B_{0}k_{0}}{\Omega_1}\right) \sin(\omega_0 t)\bigg{]},\label{MovBcomplex}\\
\!\al\al\delta_{01}\rho_{01}\frac{\partial v_{\pm 1}}{\partial t}\pm
i\rho_{02}\Omega_{2}(v_{\pm 1}-v_{\pm 2})-i \frac{k B_{0}}{\mu_{0}}B_{\pm}\nonumber\\
\al=\al\var \Big{[}-\delta_{01}\rho_{01}k_{0}v_{x1}'\left(
A_{x1}\sin(\omega_{0}t)\pm i A_{y1}\cos(\omega_0 t)\right)\nonumber\\
\al\mp\al i \Omega_{2}\rho_{02}\frac{k_0}{\omega_0}A_{x2}\cos(\omega_{0}t)(v_{\pm 1}-v_{\pm 2})\Big{]}.
\label{Movcomplex}\end{eqnarray}
Let us define the sense of polarization in reference to the screw sense of the
fields in the direction of propagation in the $z$-direction.
Then $v_+=v'_x+i v'_y$ corresponds
to a left hand circularly polarized wave for positive frequencies
and a right hand circularly polarized wave for negative
frequencies, while $v_-=v'_x-i v'_y$ corresponds to a right hand
circularly polarized wave for positive frequencies and a left hand
circularly polarized wave for negative frequencies.
Also, $B_{\pm}$ is defined similarly. With these definitions
$v_{x}', v_{y}', B_{x}', B_{y}'$ may be eliminated entirely from
these expressions. Note that the presence of a second moving ion species
introduces coupling through terms of the form: $v_{\pm 1}-v_{\pm 2}$,
and an additional term on the right hand side in Eq.\,(\ref{Movcomplex})
due to density variations.

In an analogous way to the classic driven pendulum problem \cite{MovPippard}, we have found that
the velocities of the two species satisfy a pair of generalized Mathieu equations.
Furthermore, the two species and magnetic field are strongly coupled.
To proceed, we move from the time domain to the frequency domain,
under the Fourier transform:
\begin{eqnarray}
V_{\pm 1}\!(\omega)\al=\al
\frac{1}{\sqrt{2\pi}}\int_{\!\!-\infty}^{\infty}\!v_{\pm 1}\!(t) e^{i\omega t}dt,\nonumber\\
V_{\pm 2}\!(\omega)\al=\al
\frac{1}{\sqrt{2\pi}}\int_{\!\!-\infty}^{\infty}\!v_{\pm 2}\!(t) e^{i\omega t}dt,
\end{eqnarray}
where $\omega$ is the frequency of the excited waves.
Using the linearity of our pair of differential equations, we compute
the Fourier transform of both. Upon eliminating $B_{\pm}$ and $V_{\pm 2}\!(\omega)$ in favor
of $V_{\pm 1}\!(\omega)$, we obtain:
\begin{widetext}
\begin{eqnarray}
F_{\pm}\!(\omega) V_{\pm 1}\!(\omega)
\al=\al\bar{\varepsilon}\Big{[} -\frac{1}{2}(b v_{A1}^2
k^2+P_{\pm}\!(\omega))\left(V_{\pm 1}\!(\omega_{+})+V_{\pm 1}\!(\omega_{-})\right)
+\frac{1}{2}P_{\pm}\!(\omega)\left(V_{\pm 2}\!(\omega_{+})+V_{\pm 2}\!(\omega_{-})\right)
\nonumber\\
\al+\al M_{\pm}\!(-A_{x1},+A_{x1},\omega_+,-1)V_{+1}\!(\omega_+)
+N_{\pm}\!(-A_{x2},+A_{x1},-1)V_{+2}\!(\omega_+)\nonumber\\
\al+\al M_{\pm}\!(+A_{x1},-A_{x1},\omega_-,-1)V_{+1}\!(\omega_-)
+N_{\pm}\!(+A_{x2},-A_{x1},-1)V_{+2}\!(\omega_-)\nonumber\\
\al+\al M_{\pm}\!(-A_{x1},-A_{x1},\omega_+,+1)V_{-1}\!(\omega_+)
+N_{\pm}\!(-A_{x2},-A_{x1},+1)V_{-2}\!(\omega_+)\nonumber\\
\al+\al M_{\pm}\!(+A_{x1},+A_{x1},\omega_-,+1)V_{-1}\!(\omega_-)
+N_{\pm}\!(+A_{x2},+A_{x1},+1)V_{-2}\!(\omega_-)\Big{]},
\label{Movfourdisp}
\end{eqnarray}
where,
\begin{eqnarray} F_{\pm}\!(\omega)
\al :=\al-\delta_{01}\omega^{2}+v_{A1}^{2}k^{2}
\left(1\mp\frac{\omega}{\Omega_{1}}\right) \pm
\frac{\Omega_{1} \delta_{01}} {\delta_{02}}\omega
\left( 1-\frac{1\mp\frac{\omega}{\Omega_{1}}}
{1\mp\frac{\omega}{\Omega_{2}}}\right),\label{MovFeqn}\\
P_{\pm}\!(\omega)\al :=\al
\pm\frac{\Omega_{1}\delta_{01}}{\delta_{02}}\omega
\left(\frac{k_{0}V_{x2}}
{\omega_{0}}-\frac{b}{1\mp\frac{\omega}{\Omega_{2}}}\right),\\
\!\!\!\!\! M_{\pm}\!(A_1,D_1,\xi,\epsilon)
\al:=\al-\frac{k_0}{4}\bigg{[}(\delta_{01}\omega\pm\frac{v_{A1}^2
k^2}{\Omega_i}) (A_1\pm A_{y1})
\pm(-\delta_{01}\xi+\epsilon(1-\delta_{01})\Omega_1)\left(1-\frac{1}{\delta_{02}
(1\mp\frac{\omega}{\Omega_{2}})}\right)(D_1+A_{y1})\bigg{]},\,\,\,\,\,\,
\label{MovHdef}\\
N_{\pm}\!(A_2,D_1,\epsilon)\al:=\al\pm\frac{k_0}{4}\epsilon(1-\delta_{01})\Omega_{1}
(D_1+A_{y1})
-\frac{k_0}{4}\frac{\Omega_1\delta_{01}\omega}{\delta_{02}\Omega_2}
\frac{A_2\pm A_{y2}}{1\mp\frac{\omega}{\Omega_2}},
\end{eqnarray}
\end{widetext}
and $\omega_{\pm}:=\omega\pm\omega_0$.
Eq.\,(\ref{Movfourdisp}) is obviously not an algebraic expression from which we can uniquely
obtain the dispersion equation between $\omega$ and $k$.
Instead it provides us with a functional relationship between the
Fourier transforms with arguments $\omega$, $\omega-\omega_{0}$ and $\omega+\omega_{0}$.

Nevertheless, we may proceed by noting that the right hand is $O(\bar{\varepsilon})$,
which is small. For the left hand side to be small we must ensure that the frequency
is near a root of the polynomial on this side with a correction of $O(\var)$, i.e.,
\begin{equation} \omega=\chi+O(\bar{\varepsilon}),
\label{Movnroot}\end{equation}
where $\chi$ is defined to satisfy,
\begin{equation}
F_{\pm}\!(\chi)=0.
\label{Movcubic}\end{equation}
These define the naturally occurring modes in the absence
of pumping, (i.e., if $\bar{\varepsilon}=0$ then $\omega=\chi$).
The solutions of these algebraic expressions give the dispersion relations characterizing plane transverse
waves propagating parallel to the background magnetic field in a two species plasma.
The last term of $F_{\pm}$ is only present when
a secondary moving ion-like species resides in the plasma. It is an additional Hall-type term,
associated with the relative motion of the two charged species.
In the one species limit this relation simplifies to
\begin{equation}
-\chi^{2}+v_{A1}^{2}k^{2} \left(1\mp\frac{\chi}
{\Omega_{1}}\right)=0,
\label{Movquad}\end{equation}
and we are left with the familiar fast and slow (ion-cyclotron) transverse waves \cite{MovCramer}.

The important feature of the two-species result in Eq.\,(\ref{Movcubic})
is that it is a cubic in $\chi$ and hence another
mode of excitation has been added.
Moreover, since we may choose $F_{+}$ or $F_{-}$ it follows that
there are a total of six solutions; three are
left hand circularly polarized and
three are right hand circularly polarized.
However, only three of these are physically different.
This is because if $\chi$ satisfies $F_{+}(\chi)=0$, then
$-\chi$ satisfies $F_{-}(-\chi)=0$, and vice versa.
We may then concentrate on the positive frequency solutions.

\begin{figure}
\centerline{\includegraphics[width=10cm]{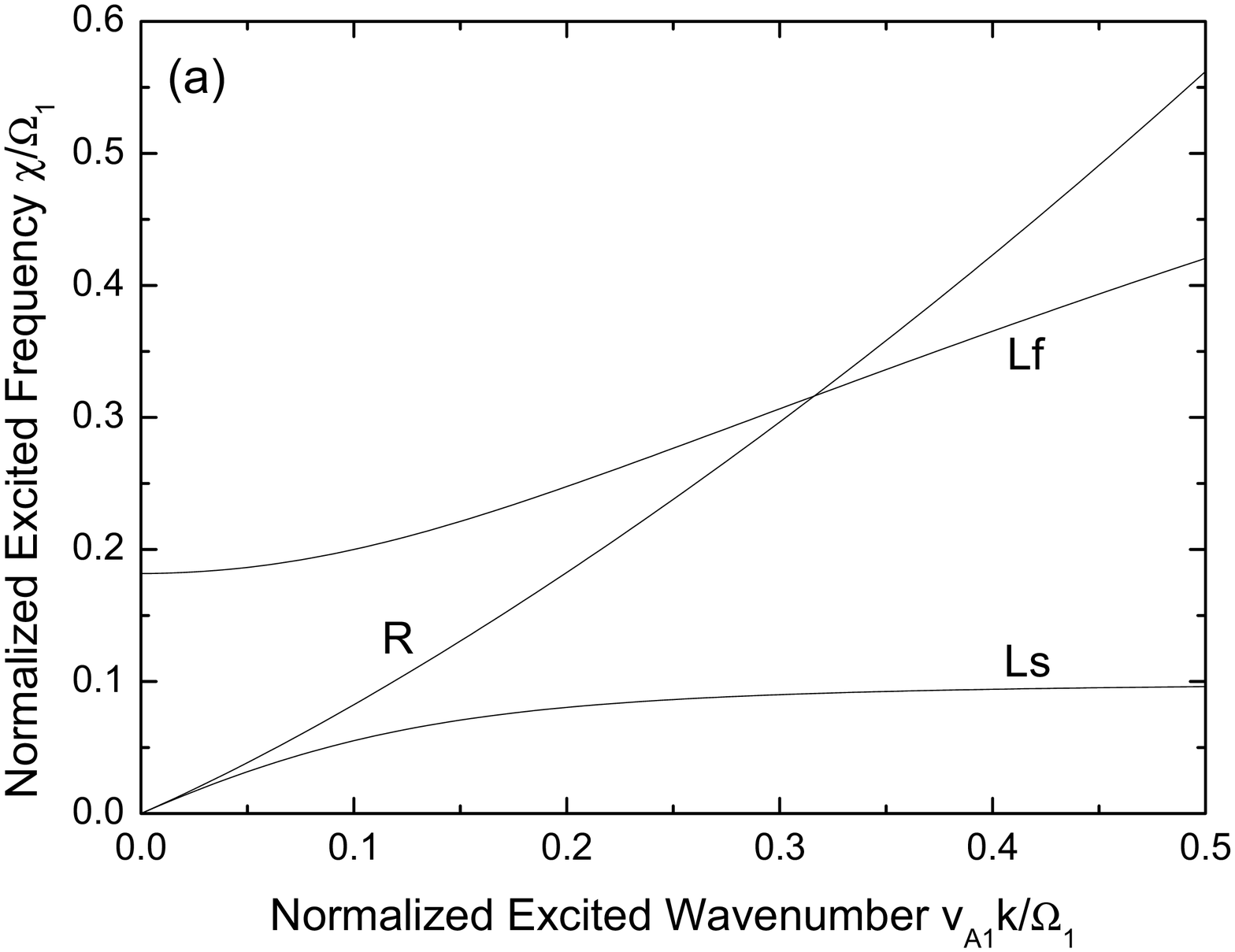}}
\centerline{\includegraphics[width=10cm]{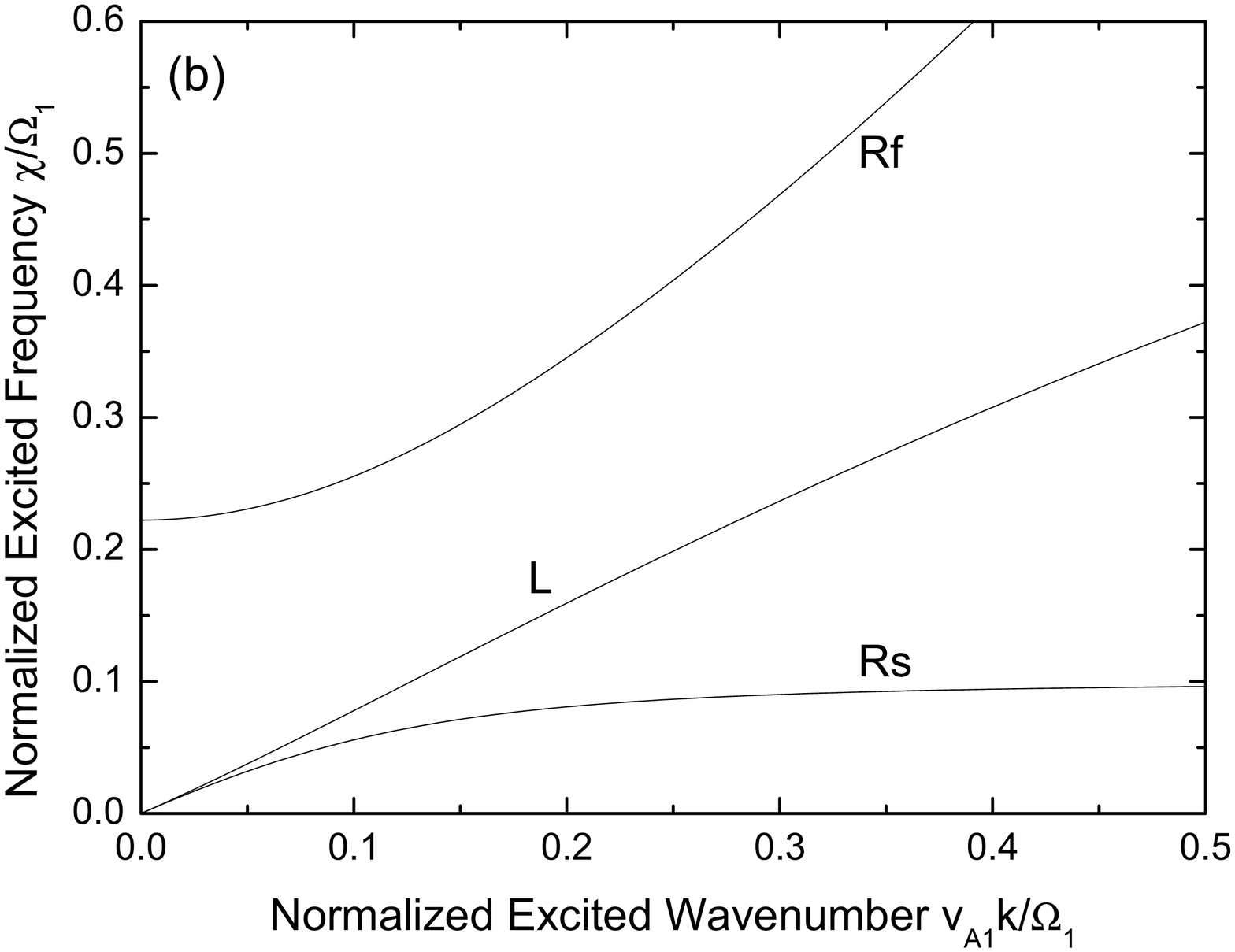}}
  \caption{The normalized
excited wave frequencies $\chi/\Omega_1$ versus normalized
wavenumber $v_{A1} k/\Omega_1$ of the three natural modes.
(a) The second species is positively charged, with $\delta_{01}=1.1$ and $\Omega_{2}/\Omega_1=0.1$.
(b) The second species is negatively charged, with $\delta_{01}=0.9$ and $\Omega_{2}/\Omega_1=-0.1$.}
\label{Mov4}\end{figure}

If the secondary species is positively
charged then there are two left hand modes of excitation \cite{MovCrambk}.
These we denote by $\omega_{Ls}$ and $\omega_{Lf}$,
where the ``{\it s}" denotes slow, and the ``{\it f}" denotes fast.
There is also a single right hand mode, denoted by $\omega_{R}$.
The corresponding dispersion relations are plotted in  Fig.\,\ref{Mov4}(a).
Note that the right hand mode intersects with the fast
left hand mode. Also, the two left hand modes both experience
a resonance as $k\to\infty$. For $\omega_{Lf}$ it is
given by $\Omega_{1}$, and for $\omega_{Ls}$ it is given by
$\Omega_{2}$ (where we have assumed $\Omega_{1}>\Omega_{2}$).

For a negatively charged secondary species
there are also three modes of excitation, however
the combination of polarizations has changed. In this case
there are two right hand modes $\omega_{Rf}$ and
$\omega_{Rs}$, and a single left hand mode $\omega_{L}$;
see Fig.\,\ref{Mov4}(b). Here $\omega_{Rs}$ has a resonance at $-\Omega_{2}$, and $\omega_{L}$
has a resonance at $\Omega_{1}$.
In the dusty plasma case, further properties of these modes were investigated
in Refs.\,\cite{MovVlad,MovMendis,MovMeuris}.

Note that the upper curves (fast modes), regardless of the sign of the species,
experience a nonzero cutoff frequency which coincides with that of the pump frequency, see Eq.\,(\ref{Movomega_c}).

\section{Growth Rates of Parametric Interaction}\label{MovResonance}
\begin{figure}
\centerline{\includegraphics[width=10cm]{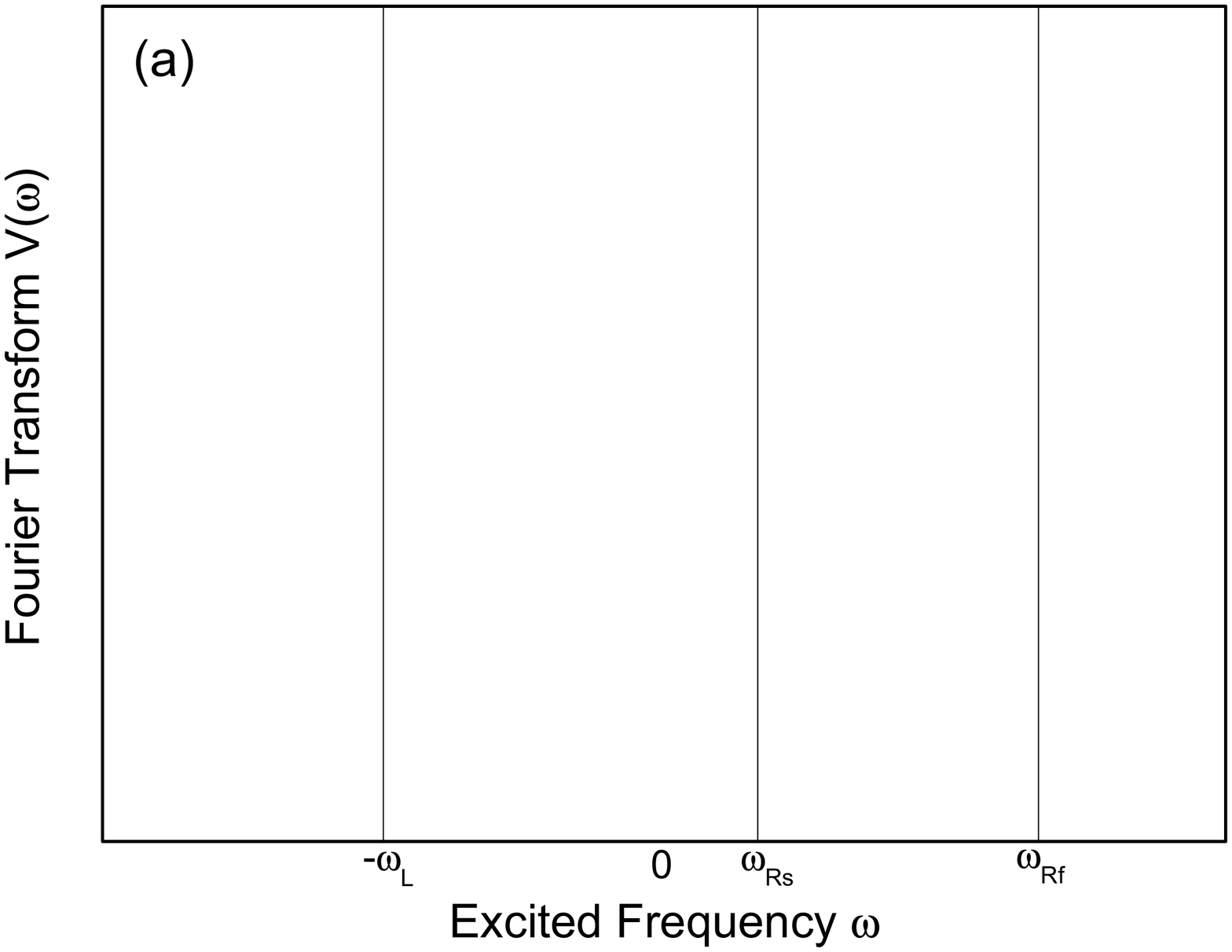}}
\centerline{\includegraphics[width=10cm]{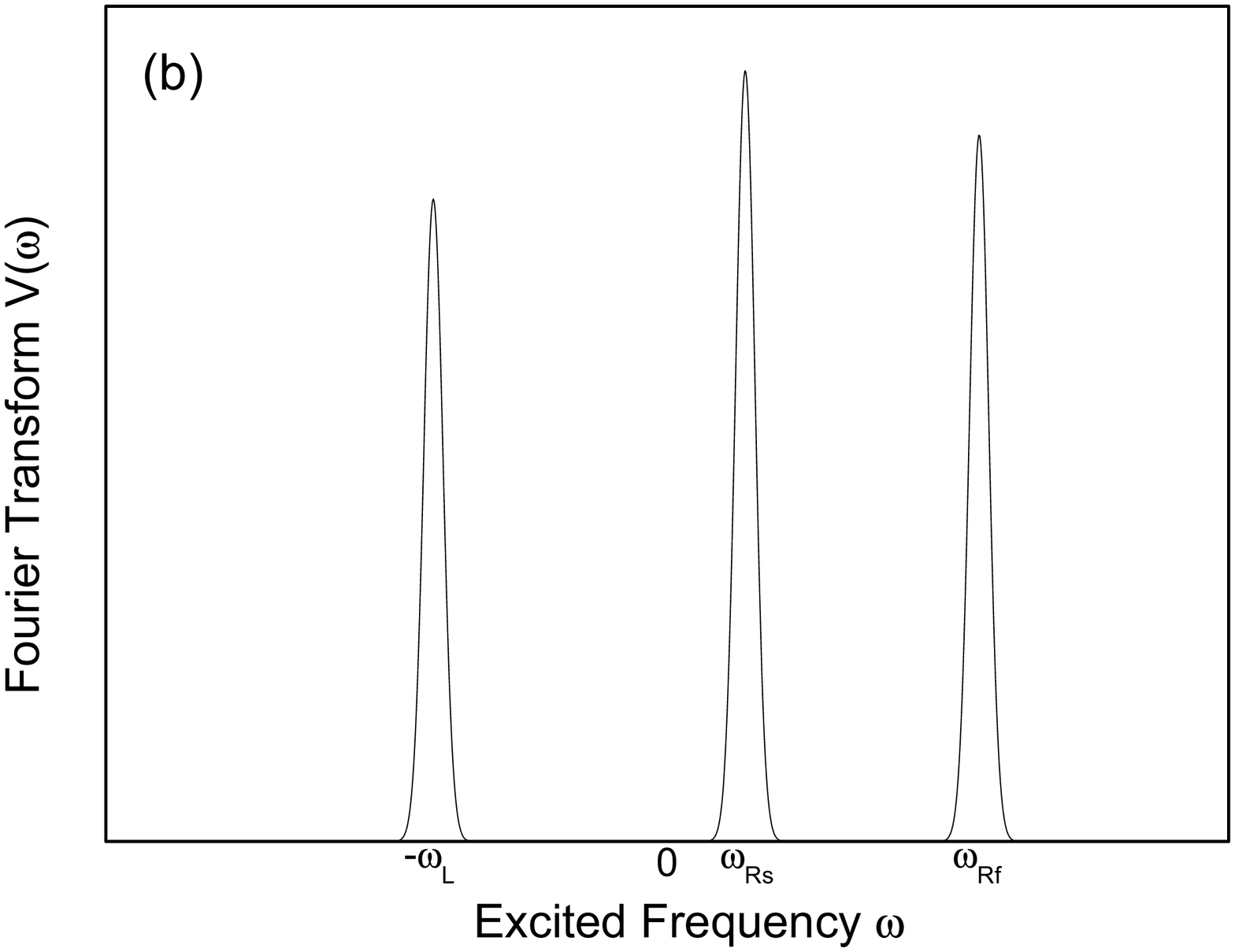}}
  \caption{A representation of the Fourier transform $V\!(\omega)$
versus frequency $\omega$.
(a) No pump, giving three Dirac-delta functions.
(b) The effect of a pump, broadening the spectrum.
Note that we have chosen to illustrate the idea using
only the right hand modes (i.e., $V_{-}$),
for a plasma with a negatively charged secondary species.}\label{Mov5}
\end{figure}

From the above analysis we see that $\chi$ may be any one of $\omega_{Ls}$,
$\omega_{Lf}$, $\omega_{R}$ for a positively charged secondary
species, or $\omega_{Rs}$, $\omega_{Rf}$, $\omega_{L}$ for
a negatively charged secondary species.
Since the frequency describing the excited wave is a perturbation of
a natural mode frequency, we denote the change by $\phi$ which is allowed to be complex, where
\begin{equation}
\omega=\chi+\bar{\varepsilon}\phi .
\label{Movcorrection}\end{equation}
Now by returning to the Fourier transform relationship in
Eq.\,(\ref{Movfourdisp}) we can see what effect $\phi$ (i.e., pumping)
will have on $V_{\pm}$. Without any pumping (i.e.,
$\bar{\varepsilon}=0$) the spatial solutions would be a linear
superposition of three pure monochromatic exponentials. Hence the Fourier transform
would be a sum of three Dirac-$\delta$ functions; see Fig.\,\ref{Mov5}(a) for a representation.
The effect of pumping is to modify this and provide frequency shifts. The modified
solution will have some broadened frequency spectrum, since it is the case that Dirac-$\delta$
functions do not solve the Fourier transformed equations in the presence of a pump.
These types of modifications will occur near each natural frequency;
see Fig.\,5(b) for a simplified representation of this effect. In fact an exact solution would be more
complicated with the identification of harmonics etc, but this figure should illuminate the key
feature of the interaction.

This is described as a {\it parametric interaction}
between the excited fields due to the pump fields.
In order to proceed we consider the case in which the interaction is greatest,
as in Refs.\,\cite{MovVahal,MovCramer}.
It is clear that this effect will be greatest when the
right hand side of Eq.\,(\ref{Movfourdisp}) is large, which
is the resonance condition. This occurs when one of the Fourier transforms on the right hand side
is resonant, in other words when their arguments are near one of the $\chi$'s, as conveyed in Fig.\,\ref{Mov5}.
Since the right hand side of Eq.\,(\ref{Movfourdisp}) involves terms
of the form: $V\!(\omega_+)$ and $V\!(\omega_-)$,
it follows that there will be a large parametric interaction when
$\omega+\omega_{0}$ {\em or} $\omega-\omega_{0}$ satisfies this.
Now, as mentioned earlier we are restricting our attention
to positive $\omega$ (without loss of generality).
Under this condition, it is found that only $\omega-\omega_{0}$ can satisfy this.
Moreover, there are three choices for this interaction:
a left hand mode interacting with a right hand mode,
a left hand mode interacting with a left hand mode, or
a right hand mode interacting with a right hand mode.

To illustrate, suppose the secondary species is
negatively charged and $\omega\approx\omega_{L}$ (i.e.,
$\chi=\omega_{L}$). Then we can have
$\omega-\omega_{0}\approx-\omega_{Rs}$ or
$\omega-\omega_{0}\approx-\omega_{Rf}$ (left hand mode
interacting with right hand mode) or
$\omega-\omega_{0}\approx-\omega_L$ (left hand mode interacting with
left hand mode). In other words,
the resonance condition is that the pump frequency satisfies
\begin{equation}
\omega_{0}=\omega_{L}+\omega_{Rs},\,\,
\omega_{0}=\omega_{L}+\omega_{Rf},\,\,\mbox{or}\,\,
\omega_{0}=2\omega_{L},
\end{equation}
respectively, which is a statement of conservation
of energy. That is, the resonance condition describes the coupling of
a pump wave with two daughter waves.

Moreover, if $\omega\approx\omega_{Rs}$ then
we can have $\omega-\omega_{0}\approx-\omega_{L}$, $\omega-\omega_0\approx-\omega_{Rf}$
or $\omega-\omega_0\approx-\omega_{Rs}$ and
the same conditions for resonance apply. Similar rules apply
for $\omega\approx\omega_{Rf}$ and for a plasma
where the secondary species is positive.

Let us now address the issue of the treatment of the wavenumber $k$ in our pair
of interacting waves. First, note that in the $z$-direction the {\it pump} wavenumber $k_0$ is zero.
For the natural modes we have that
a wave given by $(k,-\chi)$ is physically equivalent to
a wave given by $(-k,\chi)$. Thus our two interacting waves
will be given by $(k,\chi_{1})$ and $(-k,\chi_{2})$,
where $\chi_1, \chi_2$ are any combination of
$\omega_{L}, \omega_{Ls}, \omega_{Lf}, \omega_{R}, \omega_{Rs}, \omega_{Rf}$.
Hence the wavenumbers associated with each interacting wave
are equal and opposite, i.e.,
\begin{equation}
k_{1}+k_{2}=0,
\end{equation}
which is a statement of conservation of momentum in the $z$-direction.
The fact that the right hand side is zero reflects the spatial uniformity of the pump wave.
This tells us that if the approximately spatially uniform standing pump wave decays, then it does
so into two daughter waves of equal wavelength travelling in opposite directions.
Since the wavenumber magnitudes are equal we may just consider the wavenumber $k$
without referring to the sense of polarization.
This is a direct generalization of the previous investigations in \cite{MovVahal,MovCramer,MovCrambk}.

With the above framework we now proceed to solve for $\phi$ from the Fourier transform relationship.
We discuss the method involved in obtaining $\phi$ in
the case where a left handed wave interacts with a right handed wave, denoted $\phi_{LR}$.
Here we may assume, without loss of generality, that $\chi=\chi_L$
in Eq.\,(\ref{Movcorrection}), where $\chi_L$ is any one of $\omega_{L}, \omega_{Ls},
\omega_{Lf}$. This will interact with $-\chi_{R}$
(i.e., $\chi-\omega_{0}$ may be any one of $-\omega_{Rs},
-\omega_{Rf}, -\omega_{R}$).

Following the methodology of Ref.\,\cite{MovNishikawa}
we form another equation corresponding to (\ref{Movfourdisp}) under the
transformation $\omega\to\omega-\omega_{0}$.
In the resulting two equations, we neglect the obviously nonresonant terms $V\!(\omega_+)$ and
$V\!(\omega-2\omega_{0})$. Additionally, due to our particular
choice of interaction, i.e., left-right, we neglect $V_{+}(\omega_-)$ and $V_{-}(\omega)$,
and retain only $V_{+}\!(\omega)$ and $V_{+}\!(\omega_-)$.
We find the following:
\begin{eqnarray}
\al\al\!\!\!\! F_{+}\!(\chi_{L}+\bar{\varepsilon}\phi_{LR})
F_{-}\!(\chi_{R}-\bar{\varepsilon}\phi_{LR})\nonumber\\
\al=\al
\bar{\varepsilon}^2\left[Y_{+}\!(-\chi_{R})+J_{+}\!(+A_{x1},+A_{x2},-A_{x1},-\chi_{R},-1)\right]\nonumber\\
\al\al\times\,     \left[Y_{+}\!(+\chi_{L})+J_{+}\!(-A_{x1},-A_{x2},+A_{x1},+\chi_{L},-1)\right],\,\,\,\,\,\,\,\,\,
\label{Movmaindisp}\end{eqnarray}
where
\begin{eqnarray}
\al\al Y_{\pm}\!(\xi) :=
-\frac{1}{2}\left(b v_{A1}^{2}k^{2}+P_{\pm}\!(\omega)\left(1-\frac{1\pm\frac{\xi}{\Omega_{1}}}
{1\pm\frac{\xi}{\Omega_{2}}}\right)\right),\,\,\,\\
\al\al J_{\pm}\!(A_1,A_2,D_1,\xi,\epsilon) :=
\frac{1\mp\frac{\xi}{\Omega_1}}{1\mp\frac{\xi}{\Omega_2}} N_{\pm}\!(A_2,D_1,\epsilon)\nonumber\\
\al\al\,\,\,\,\,\,\,\,\,\,\,\,\,\,\,\,\,\,\,\,\,\,\,\,\,\,\,\,\,\,\,\,\,\,\,\,\,\,\,\,\,\,\,\,\,
\,\,\,\,\,\,\,\,\,\,\,\,\,\,\,\,\,\,\,\,\,\,\,\,\,\,\,\,\,\,
+M_{\pm}\!(A_1,D_1,\xi,\epsilon).
\end{eqnarray}
In these expressions we implicitly have that $\xi$ and $\omega$,
are related by: $\xi\to\chi_L\Rightarrow\omega\to-\chi_{R}$
and $\xi\to\chi_R\Rightarrow\omega\to-\chi_{L}$.

In the one ion species analysis, it is at this point that
the conservation of energy and momentum rules are used to explicitly obtain
$\chi_{L}, \chi_{R}, k$. The procedure is to use
$F_{\pm}$ to solve for $\chi_{L,R}$ in terms of $k$.
These solutions are then added together and equated to $\omega_{0}$.
The resultant expression
is then solved for $k$ and subsequently substituted into
the expressions for $\chi_{L,R}$. This process
explicitly shows that the resonance condition,
for a given value of pump frequency $\omega_{0}$,
uniquely determines the frequencies and wavenumbers of the excited waves.
Now, this is all possible since the solution of $F_{\pm}$
in the one species limit is merely a quadratic, see
Eq.\,(\ref{Movquad}). However, in our case
we must solve a cubic. As such, the resulting expressions
for $\chi_{L}, \chi_{R}, k$ are too complicated to be
reproduced here.  Let us just note that it is still
true that $\omega_{0}$ determines them uniquely.

In order to solve for $\phi_{LR}$ in Eq.\,(\ref{Movmaindisp})
we Taylor expand $F_{+}\!(\chi_{L}+\bar{\varepsilon}\phi_{LR})$
around $\chi_{L}$ to order $\bar{\varepsilon}$, then use the fact that
$F_{+}\!(\chi_{L})=0$, to obtain
\begin{equation}
F_{+}\!(\chi_{L}+\bar{\varepsilon}\phi_{LR})=
\bar{\varepsilon}\phi_{LR} H\!(\chi_{L}),
\end{equation}
where
\begin{equation}
H\!(x):=-\frac{v_{A1}^{2} k^2}{\Omega_{1}}-\delta_{01}
x\left(2+\frac{(\Omega_{1}-\Omega_{2})(2\Omega_{2}-x)}
{\delta_{02}(\Omega_{2}-x)^{2}}\right),
\end{equation}
while $F_{-}\!(\chi_{R}-\bar{\varepsilon}\phi_{LR})$ can be treated
similarly. Hence $\phi_{LR}$ is obtained from:
\begin{eqnarray}
\phi_{LR}^{2}\!\al=\al\!\left[Y_{+}\!(-\chi_{R})+J_{+}\!(+A_{x1},+A_{x2},-A_{x1},-\chi_{R},-1)\right]\nonumber\\
\al\,\times\al\left[Y_{+}\!(+\chi_{L})+J_{+}\!(-A_{x1},-A_{x2},+A_{x1},+\chi_{L},-1)\right]\nonumber\\
\al\times\al\left[H\!(\chi_{L})\,H\!(-\chi_{R})\right]^{-1}.\,\,\,\,\,\,\,\,\,\,\,
\end{eqnarray}

Although this was derived assuming we were near a natural
frequency of a left hand mode, we obtain the same
expression for $\phi^2$ for the right hand case (viz. $\phi_{RL}^2$)
with the modification that $\phi_{RL}$ is of the opposite sign.

This technique may also be applied to both the left-left
interactions and the right-right interactions. This includes the possibility of a
slow wave interacting with a fast wave (e.g., $\omega_{Ls}$ and $\omega_{Lf}$),
and waves of equal frequency interacting.
These sorts of interactions occur only because of the velocity derivative terms
in (\ref{Movvelderiv}) \& (\ref{Movvelderiv2}).
In obtaining $\phi$ we must retain and neglect the appropriate selection of Fourier
transforms according to the choice made. Here we introduce $\chi_{L1}$ and $\chi_{L2}$
as two (possible equal) left hand modes, and $\chi_{R1}$ and $\chi_{R2}$
as two (possible equal) right hand modes.
The corresponding frequency changes are obtained from
\begin{eqnarray}
\phi_{LL}^2\al=\al
                    \left[J_{+}\!(+A_{x1},+A_{x2},+A_{x1},-\chi_{L2},-1)\right]\nonumber\\
\al\,\,\,\,\times\al\left[J_{-}\!(-A_{x1},-A_{x2},+A_{x1},+\chi_{L1},-1)\right]\nonumber\\
\al\,\,\,\,\times\al\left[H\!(\chi_{L1})\,H\!(-\chi_{L2})\right]^{-1},\\
\phi_{RR}^2\al=\al
                    \left[J_{-}\!(+A_{x1},+A_{x2},-A_{x1},-\chi_{R2},-1)\right]\nonumber\\
\al\,\,\,\,\times\al\left[J_{+}\!(-A_{x1},-A_{x2},-A_{x1},+\chi_{R1},+1)\right]\nonumber\\
\al\,\,\,\,\times\al\left[H\!(\chi_{R1})\,H\!(-\chi_{R2})\right]^{-1},
\end{eqnarray}
for the left-left and right-right interactions, respectively.

It is worth noting that in other types of parametric problems, such as modulational and beam instabilities, a 
graphical approach is useful in classifying the instabilities; they occur where two normal mode lines cross and 
reconnect \cite{MovLongtin,MovGomberoff}. In those cases the number of interacting modes is finite, and the nonlinear 
dispersion relation can be obtained in closed form. In our case, however, we cannot obtain the nonlinear dispersion 
relation in closed form, due to the infinite number of interacting modes; the decay interaction of the two excited 
modes with the pump is postulated in the first approximation, and the classification of the instability is relatively 
straightforward.

\section{Numerical Analysis}
From the expressions for $\phi_{LR}^2, \phi_{LL}^2, \phi_{RR}^2$ it is easy to see
that these quantities are real valued when we are in a frequency-wavenumber
regime such that a pump wave can propagate. Hence $\phi$ is either purely real
or purely imaginary. If $\phi$ is real then what we have found
is a correction to the natural mode frequency which adds an
extra Fourier component to the field expansions.
If $\phi$ is imaginary then the daughter waves will either undergo
exponential growth or decay. Moreover, since there will exist
both a positive and a negative imaginary frequency solution and
since the frequency is conserved, this implies that
one of the daughter waves will grow and the other will decay.
In such cases the presence of one exponentially growing daughter wave implies
that anywhere in which $\phi^{2}<0$ is a region of instability.
In this case $\phi$ is the growth rate.

\begin{figure}
\centerline{\includegraphics[width=10cm]{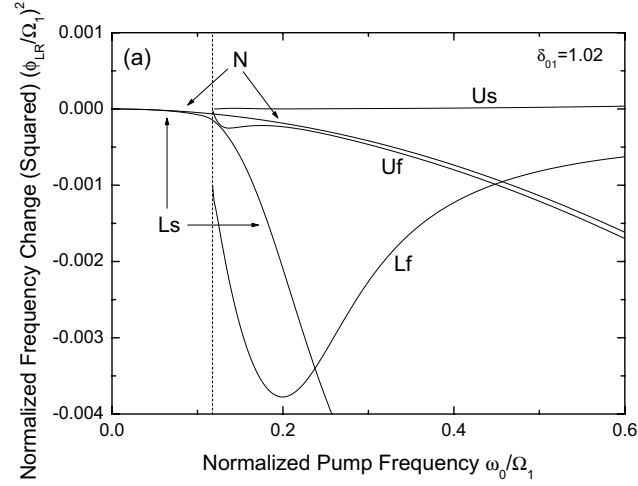}}
\centerline{\includegraphics[width=10cm]{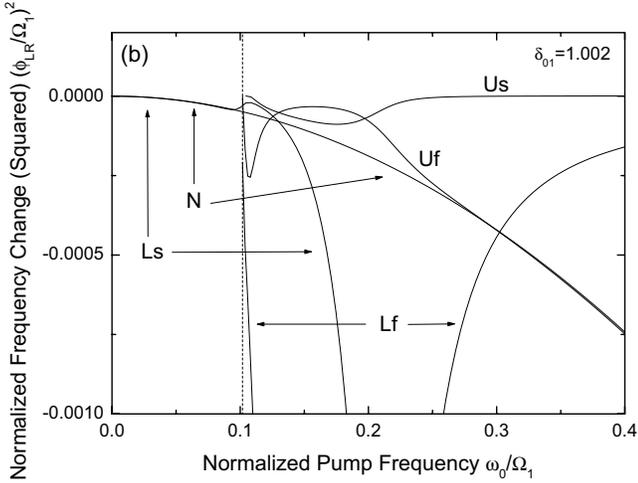}}
  \caption{The squared normalized frequency change
 $(\phi_{LR}/\Omega_{1})^{2}$ for the left-right interaction plotted against
 pump frequency $\omega_{0}/\Omega_{1}$.
In both plots the second species is positive, with
$\beta_{1}=0.8,\, B_{1}=1.5,\, U_{2}/U_{1}=1/2$ and $\Omega_2/\Omega_1=0.1$.
 The number densities vary in (a)--(b), with
 $\delta_{0 1}=1.02$ in (a), and $\delta_{0 1}=1.002$ in (b).
 Also, ``$Us$, $Uf$, $Ls$, $Lf$" denotes the upper-slow,
 lower-slow, upper-fast and lower-fast combinations,
 respectively. Here ``$N$" denotes the single species result.}
\label{Mov6}\end{figure}

In Fig.\,\ref{Mov6} we plot the full complement of (normalized)
squared frequency changes.
That is, we plot $(\phi_{LR}/\Omega_{1})^{2}$ versus pump frequency
$\omega_{0}/\Omega_{1}$ for the upper ``$U$" and lower ``$L$" branches of the pump
dispersion relation for each mode of interaction.
This figure is for a warm plasma in which the second heavy species
is positively charged. The same basic features are present in
the negatively charged case. All curves labelled ``$s$", i.e., ``$Us$" and ``$Ls$",
correspond to the combination between $\omega_{R}$ and $\omega_{Ls}$,
that is the {\it slow} interaction. The curves labelled ``$f$", i.e., ``$Uf$" and ``$Lf$",
are for the {\it fast} interaction between $\omega_{R}$ and $\omega_{Lf}$.

In Fig.\,\ref{Mov6}(a), where $\delta_{01}=1.02$,
the full spectrum of perturbations to the natural modes is seen.
The $Uf$ interaction is unstable and monotonically decreases,
approaching the single species result $N$ for large $\omega_{0}$.
Hence this is indeed consistent with the single ion analysis, given by
\begin{equation}
\phi_{LR}^2\to -\frac{v_{A}^4 k^4}{4\omega_{0}^2(1+\beta_1+B_1)},
\,\,\,\, \mbox{as}\,\,\,\, \delta_{01}\to 1.
\end{equation}

The $Lf$ interaction is also unstable, however it experiences a
minimum (corresponding to a {\it maximum} growth rate), about which it turns over and approaches zero. For our
choice of parameters, in this figure, this occurs near
$\omega_{0}\sim 0.2\,\Omega_{1}$. Moreover the intersection with $N$ occurs
at $\omega_0\sim 0.45\,\Omega_1$. Next, we see that the $Us$ interaction,
is small and positive, and hence it is weakly stable. Note, however,
that if higher order terms are included in the field expansions
then this may in fact prove to be weakly unstable.
Finally, we see that the $Ls$ interaction, which is only present
in a {\it warm} plasma, is strongly unstable, with a growth rate that approaches infinity at
a rate considerably faster than $N$ or $Uf$.

As we decrease the number density of the second species in
Fig.\,\ref{Mov6}(b) to $\delta_{01}=1.002$ some interesting features appear.
The $Uf$ curve shows a sharply varying growth rate at just above the cutoff frequency.
The $Lf$ minimum shifts to a lower value (lying beyond the range of the plot).
The $Us$ interaction has actually become
unstable in the frequency region just above the cutoff frequency.
Also, the manner in which our bi-ion result approaches the
single species result as $\delta_{01}\to 1$ is interesting:
the envelope of the three modes $Ls$, $Us$, and $Uf$
form $N$ in the domains $\omega_{0}\lesssim\omega_{c}$,
$\omega_{c}\lesssim\omega_{0}\lesssim 2\,\omega_{c}$ and $\omega_{0}\gtrsim 2\,\omega_{c}$, respectively.

\begin{figure}
\centerline{\includegraphics[width=10cm]{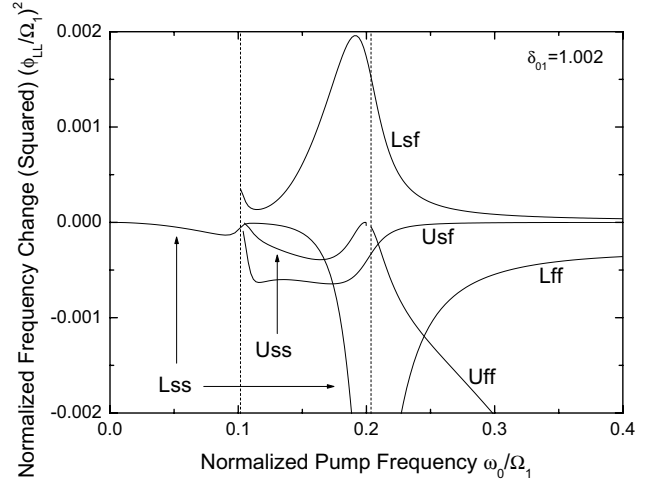}}
  \caption{The squared normalized frequency change
 $(\phi_{LL}/\Omega_{1})^{2}$ for the left-left interaction plotted against
 pump frequency $\omega_{0}/\Omega_{1}$.
Here the second species is positive, with
$\beta_{1}=0.8,\, B_{1}=1.5,\, U_{2}/U_{1}=1/2,\, \Omega_2/\Omega_1=0.1$
and $\delta_{0 1}=1.002$.
 Also, $Uss$, $Lss$, $Uff$, $Lff$, $Usf$, $Lsf$ denotes the upper-slow-slow,
 lower-slow-slow, upper-fast-fast, lower-fast-fast, upper-slow-fast and lower-slow-fast combinations,
 respectively.}\label{Mov7}
\end{figure}

In Fig.\,\ref{Mov7} we plot the range of left-left interactions for the positively charged secondary species case.
As can be seen in Fig.\,\ref{Mov4}(a) this case has two left handed branches (slow and fast). Hence
we may form the slow-slow, fast-fast and slow-fast combinations. This combined with the
lower and upper branches of the pump dispersion relation gives a total of six interactions.
Of these six we find that all are unstable, except the ``$Lsf$" combination which experiences
a point of maximum stability at just below $2\omega_{c}$.
Note that the ``$Lss$" interaction decreases without bound as $\omega_{0}\to 2\Omega_1$
This is acceptable given that for the slow-slow case the excited wavenumber $k$ experiences
a resonance as $\omega\to\Omega_1$ (with pump wavenumber small).
Although it lies below the range of the plot the ``$Lff$" interaction starts at a finite value,
when $\omega_0=2\omega_c$. Also note the curious feature wherein the ``$Usf$" interaction
has a corresponding growth rate which is large in the range $\omega_c\lesssim\omega_0\lesssim 2\,\omega_c$ only.

\begin{figure}
\centerline{\includegraphics[width=10cm]{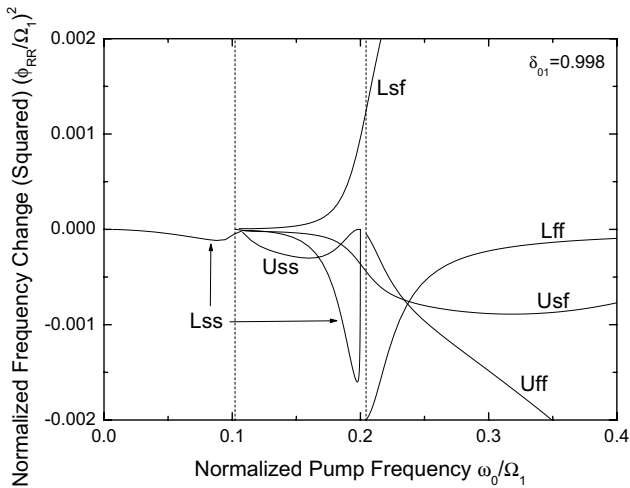}}
  \caption{The squared normalized frequency change
 $(\phi_{RR}/\Omega_{1})^{2}$ for the right-right interaction plotted against
 pump frequency $\omega_{0}/\Omega_{1}$.
Here the second species is negative, with
$\beta_{1}=0.8,\, B_{1}=1.5,\, U_{2}/U_{1}=1/2,\, \Omega_2/\Omega_1=-0.1$
and $\delta_{0 1}=0.998$.
 Also, ``$Uss$, $Lss$, $Uff$, $Lff$, $Usf$, $Lsf$" denotes the upper-slow-slow,
 lower-slow-slow, upper-fast-fast, lower-fast-fast, upper-slow-fast and lower-slow-fast combinations,
 respectively.}\label{Mov8}
\end{figure}

Finally, in Fig.\,\ref{Mov8} we plot the range of right-right interactions for the negatively
charged secondary species case.
Again by inspection of Fig.\,\ref{Mov4}(b)
it makes sense that there are six values of $\phi_{RR}^2$.
This plot shares many features in common with the previous plot.
However, an important change is that the $Lss$ interactions has a minimum and approaches zero
as $k\to\infty$. Also, the $Usf$ maximal growth rate has shifted to much higher values of pump frequency.
The $Lsf$ interaction has stability as $\omega_0$ increases.

Hence, we have seen that the introduction of a second species adds
a range of extra behaviors, even with a small number density.
Note that in the special case where the frequencies of different modes coincide,
the selection of resonant Fourier transforms to generate these plots is actually invalid.
In particular, this occurs at the point seen in Fig.\,\ref{Mov4}(b) where $\omega_{R}=\omega_{Lf}$.
At that point there are actually more terms resonant than have been accounted for.
However, since this occurs on a set of measure zero, it has been ignored.

\section{Conclusions}
This paper has outlined an investigation into a certain class of instabilities in
a bi-ion or dusty plasma, allowing full mobility of all species and the inclusion of pressure. The results should be 
applicable to a wide range of laboratory and space plasmas where secondary ion species or charged dust grains occur.
The immediate consequence of the presence of an extra heavy species is its increase or reduction of the
number of free electrons in the plasma through the charge neutrality condition.
We obtained a pump wave with spatial variation transverse to the
background magnetic field, which contained an upper and lower branch,
and made the approximation of spatial uniformity of this pump.
The stability of this structure was investigated by perturbing the pump wave,
as we looked for parametric instabilities wherein the pump wave generates excited waves.
Three natural transverse modes were found to be excited propagating parallel to the background magnetic
field; two left handed when the second species is positive and two right handed when the second species
is negative. Two out of each of these three are the modified fast and slow (ion-cyclotron) modes.
There were three basic types of interactions allowed;
left-right, left-left and right-right, each of which has several combinations
dependent on both the choice of branch from the pump wave (upper and lower) and the choice of
natural mode (slow and fast), giving rise to twelve interactions in total for any particular plasma.

We derived the frequency changes to the natural modes, corresponding to growth rates when
the perturbation to the frequency was imaginary. We can summarize the main features of the instabilities for the 
left--right interactions (opposite sense circular polarization) as follows:

(i) Of the four combinations allowed, three are unstable when the number density
of the second massive species is substantial, and all four are unstable when the
number density of the second massive species is low.

(ii) As the single species case is approached ($\delta_{1}\to 1$),
the envelope of three different growth rates approaches it in the domains
$\omega_0\lesssim\omega_c,\,\omega_c\lesssim\omega_0\lesssim 2\,\omega_c$ and $\omega_0\gtrsim 2\,\omega_c$.

(iii) The two interactions corresponding to the lower branch of the pump dispersion
relation are strongly unstable, with the $Lf$ experiencing a local maximum and $Ls$ unbounded.

Next we investigated the left-left and right-right interactions (same sense circular polarization):

(i) These interactions involve six and two
combinations respectively when the second species is positive, and the reverse when the
second species is negative. This follows simply from the assortment of handedness of natural
modes.

(ii) Of these only the $Lsf$ interaction is stable, while the others display interesting
behaviors. In particular, in the positive case, the $Lss$ curve approaches $-\infty$ as $\omega_0\to 2\omega_c$
where the excited wavenumber $k$ experiences a resonance, while its growth rate is maximal
at $\omega_0$ just below $2|\omega_c|$ in the negative case.

(iii) In contrast to the left--right interactions, the upper pump mode can give the strongest instabilities. The 
$Usf$ interaction
has a large growth rate between $\omega_c$ and $2\omega_c$ in the positive second species case, and at higher
values of pump frequency in the negative second species case.

In several of the above interactions the growth rate
starts at a nonzero value. This occurs in the cases where the starting value of $\omega_0$
have a corresponding nonzero value for $k_0$, so the velocity derivative terms are
nonzero.

A possibility for further work would be to allow the excited waves to be fully oblique,
rather than simply parallel to the background magnetic field.
In doing so we would be able to remove the long pump wavelength assumption, which may
lead to interesting features.
In the single ion case this was treated in Ref.\,\cite{MovCramer76}, albeit without pressure.
It is anticipated that implementation of the accompanying additional wavenumber matching rules, in both
the $z$ and $x$ directions, would be algebraically complicated.

%


\end{document}